 \documentclass[preprint,10pt]{elsarticle} 

\usepackage[utf8]{inputenc}
\usepackage{graphicx}
\usepackage{amsmath}
\usepackage{subfig}
\usepackage{float}
\usepackage{cite}
\usepackage{amssymb}
\usepackage{mathtools}
\usepackage[symbol]{footmisc}

\usepackage[a4paper, total={6in, 8in}]{geometry}
\usepackage[toc,page]{appendix}

\usepackage[colorlinks,bookmarksopen,bookmarksnumbered,citecolor=blue,urlcolor=blue]{hyperref}
\usepackage[usenames,dvipsnames,svgnames,table]{xcolor}
\newcommand{\red}[1]{{\color{red} #1}}  
\newcommand{\blue}[1]{{\color{black} #1}} 
\newcommand{\green}[1]{{\color{black} #1}} 
 
 \usepackage{algorithm}
\usepackage{algpseudocode}

 \usepackage{todonotes}
\begin{document}

\begin{frontmatter}
\title{
A Lagrangian meshfree model for solidification of liquid thin-films}


\author[inst1,inst3]{Anand S Bharadwaj\footnote{Corresponding author: anandbharadwaj1950@gmail.com}$^,$}

\affiliation[inst1]{organization={Fraunhofer ITWM},
            addressline={Fraunhofer-Platz 1}, 
            city={Kaiserslautern},
            postcode={67663}, 
            country={Germany}}

\author[inst1]{Elisa Thiel}
\author[inst2]{Pratik Suchde\footnote{pratik.suchde@gmail.com, pratik.suchde@uni.lu}}

\affiliation[inst2]{organization={University of Luxembourg},
            addressline=	{2 avenue de l’universite}, 
            postcode={L-4365}, 
            state={Esch-sur-alzette},
            country={Luxembourg}}

\affiliation[inst3]{organization={Indian Institute of Technology Delhi},
            addressline=	{Hauz Khas}, 
            postcode={110016}, 
            state={New Delhi},
            country={India}}
            
\begin{abstract}
In this paper, a new method to model solidification of thin liquid films is proposed. \blue{This method is targeted at applications like aircraft icing and tablet coating where the formation of liquid films from impinging droplets on a surface form a critical part of the physics of the process.} The proposed model takes into account the (i) unsteadiness in temperature distribution, (ii) heat transfer at the interface between the solid and the surface, (iii) volumetric expansion/contraction and (iv) the liquid thin-film behaviour, each of which are either partly or fully ignored in existing models.  The liquid thin-film, modeled using the Discrete Droplet Method (DDM), is represented as a collection of discrete droplets that are tracked in a Lagrangian sense. The height of the liquid film is estimated as a summation of Gaussian kernel functions associated with each droplet. At each droplet location, a solid height is also computed. The evolution of the solid height is governed by the Stefan problem. The flow of the liquid thin-film is solved just as in the case of DDM, while also taking into consideration the shape of the solidified region lying beneath the droplet. The results presented in this work show the reliability of the proposed model in simulating solidification of thin-films and its applicability to complex problems such as ice-formation on aircraft wings. \blue{  The model has been verified for canonical problems that have analytical solutions. For the more complex problems of icing, the results of the model are compared with data from literature, without considering a background air flow. The comparison can be improved by coupling this model with suitable air flow solvers, as shown in the final test case. }
\end{abstract} 
\end{frontmatter}
\section{Introduction}

Liquid thin-films are commonly found in lubrication \cite{luo1996thin}, medicinal coatings \cite{overhoff2009use,jiang2012thin}, and in the automotive industry \cite{rabiei2013rainfall}, among many other applications. In several of these applications, the liquid thin-films are subject to freezing due to exposure to low temperatures. For instance, aircraft icing \cite{myers2001extension} is a very commonly observed case of freezing of water films, which forms layers of ice on the aircraft surface. This is also observed quite commonly on wind-turbines in cold weather \cite{molinder2018probabilistic}. Pharmaceutical thin-film freezing \cite{sahakijpijarn2020development} is yet another application, used in the production of pulmonary drugs, where a liquid thin-film is solidified over a cryogenic surface. Spray-coating of tablets too involves deposition and freezing of liquid drops over a substrate \cite{seo2020pharmaceutical}. Thus, the freezing of thin-films is an important phenomenon to study and numerical models assist with better understanding of the physics of the process.

Freezing of thin-films has been an important area of focus for modelling and simulation in aircraft icing research. Messinger \cite{messinger1953equilibrium} proposed one of the first icing models. Subsequently, Myers \cite{myers2001extension,myers2004mathematical,myers1999ice,myers2002flow,myers2002slowly} enhanced the model to accommodate more physics involved in the phenomenon of icing. In the last two decades, this model has been frequently used as a base for ice and water film predictions\cite{myers2004mathematical,gori2022modeling,chen2023numerical,beaugendre2003fensap}. In this model, the evolution of thickness of the ice and water film is tracked using a coupled pair of equations representing mass and energy balance in the water and ice layers. These equations are solved in an Eulerian framework over the entire surface. \blue{The model assumes that the surface has high  thermal conductivity, such that water incident on the surface immediately reaches the same temperature as the surface.} Additionally, the temperature distribution in the water and ice layers are assumed to be quasi-steady and thus, linear. \blue{ Gori et al. \cite{gori2018local} propose the use of an analytical solution to the unsteady Stefan problem. In this model, the temperature of the ice at the surface is assumed to be equal to the temperature of the surface. 
 Liu et al. \cite{liu2019three} solve the unsteady heat equation numerically using a finite-difference method in the ice layer, while making a quasi-steady assumption in the water layer.} \green{ Chauvin et al. \cite{chauvin2018implicit} proposed a three-layer icing model with an ice layer sandwiched between a running film and a static film. They assume the temperature of the running layer to be uniform, while accounting for unsteady variation of temperature in the ice layer. Myers et al. \cite{myers2007cubic} too proposed a three-layer model and solved the unsteady heat equation in each of the layers by assuming a cubic polynomial solution for the temperature in the individual layers. In the work of  Malik et al. \cite{malik2023experimental}, a mixture of ice and water is considered on a substrate and the unsteady heat equation is solved in the mixture. In this method, the interface between ice and water is captured using the enthalpy approach for the phase-change.} Impinging droplets on the thin liquid film have been handled using both Eulerian \cite{norde2014splashing,norde2018eulerian,beaugendre2003fensap} and Lagrangian approaches\cite{wang2016two,ruff1990users,villedieu2012sld}. The two approaches have their respective pros and cons. Since the Eulerian approach models the droplets as a continuous field, the resolution of individual droplets is not necessary and therefore, can be computationally efficient. However, in the Lagrangian approach, the droplet physics can be captured more accurately and phenomena like splash and rebound can be easily incorporated in the model. Some of the prominent icing codes that handle freezing of water films are LEWICE\cite{ruff1990users,wright2015recent,rocco2021super}, ONERA\cite{hedde1995onera,trontin2017description}, GlennICE\cite{wright2002user,wright2010mixed,wright2008comparison}, FENSAP-ICE\cite{beaugendre2003fensap,bourgault2000development}. Makonnen \cite{makkonen2000models} proposed an alternate model that predicts the rate of ice accretion based on collision, sticking and accretion efficiencies. These efficiencies are derived from empirical data fits. The model has been used for icing predictions on wind turbines and wind farms \cite{molinder2018probabilistic,strauss2020skill,davis2014forecast}. In recent times, machine learning-based models have gained popularity in predictions of thin-film behaviour \cite{martin2023physics} and have been also been used in icing predictions \cite{molinder2020probabilistic,kreutz2023ice}.  High fidelity approaches have also been used to model and solve freezing of impinging droplets over surfaces \cite{fukudome2021numerical,liu1993numerical,vu2018numerical}. In these appraoches, the complete Navier--Stokes equations are solved in the interior of each droplet. Such approaches capture the physics very accurately but become impractical when the number of droplets is large due to high computational costs. 

The model introduced in the present work, called Discrete Droplet Method with Solidification (DDM-S), relaxes some of the assumptions \green{common to many of the Messinger-family} of models stated above. \green{While many of these models, with a few exceptions \cite{chauvin2018implicit,malik2023experimental,myers2007cubic}, solve a quasi-steady thermal problem as a part of the solidification process,} the DDM-S model allows unsteady temperature distributions in the solid and liquid films. This is accounted for by evaluating the solid-liquid interface movement based on the solution of the unsteady energy equation in the solid and liquid regions. The present model also allows for the fluid in contact with the surface to have a different temperature than that of the surface. In other words, it does not assume a perfect thermal contact, as required by the extended Messinger's model\cite{myers2001extension}. This reflects the physics more accurately since the liquid need not necessarily attain the same temperature as the surface, especially on impingement. The temperature of the liquid changes with time, owing to the heat transfer occurring between the liquid and the surface. The DDM-S model also conveniently incorporates volumetric change based on the density difference between the solid and liquid phase. The change in height of the solid layer and that of the liquid thin-film, is scaled using the density ratios in order to account for the volumetric change. The DDM-S model does not rely on empirical relations and attempts to capture the behaviour from physical principles.  The basis of this model is the discrete droplet method (DDM) \cite{bharadwaj2022discrete} that governs the evolution of liquid thin-films. The DDM, owing to its ability to capture liquid thin-film behaviour accurately, forms a good basis for the solidification model. \blue{In the DDM, a thin-film is modelled as an aggregation of droplets on a surface and the dynamics of the thin-film are predicted based on the motion of these droplets. The model can capture the dynamics of fully developed thin-films as well as developing thin-films.} Several of the advantages of using the DDM \cite{bharadwaj2022discrete} are passed on to DDM-S. For instance, the DDM efficiently captures partially wetted surfaces since the descritization based on droplets is used only in regions where the thin-film is present and not over the entire surface, thus, reducing computational costs. 

The layout of the paper is as follows. Section Methodology describes in detail the different sub-parts of the proposed model, DDM-S. Some of the details that are not directly relevant to the model are described in appendixes. Section Results presents five test cases, which are arranged in an order of increasing complexity. \blue{The first three test cases validate specific parts of the model by comparing the numerical solutions obtained from DDM-S with analytical solutions. The last two test cases compare the numerical solution of the entire DDM-S model with numerical and experimental data from literature.} Finally, we present the conclusions. 
\section{Methodology}
In this section, we discuss in detail the proposed thin film solidification model. The section begins with a brief overview of DDM which is the existing liquid thin-film model that forms a base of the present work. We then proceed to introduce the new solidification model (DDM-S) in the following subsections.

\subsection{Terminology}
Typical droplet-based configurations for the DDM model and the DDM-S model are shown in Figs.~\ref{fig:DDM} and \ref{fig:solid_height} respectively. In this work, the words -- \textit{substrate} and \textit{surface}, are used interchangeably and denote the underlying surface in $\mathbb{R}^3$ over which liquid thin-films form. The liquid-thin film may be referred to as \textit{liquid-film} or \textit{thin-film}. The solidified region is sometimes also referred to as \textit{solid region} or \textit{solid layer}. The thickness of the film is referred as its \textit{height}. 

\subsection{Discrete Droplet Method (DDM)}\label{sec:DDM}
\begin{figure}
    \centering
    \subfloat[]{
    \includegraphics[trim=5cm 5cm 5cm 5cm,width=0.5\textwidth]{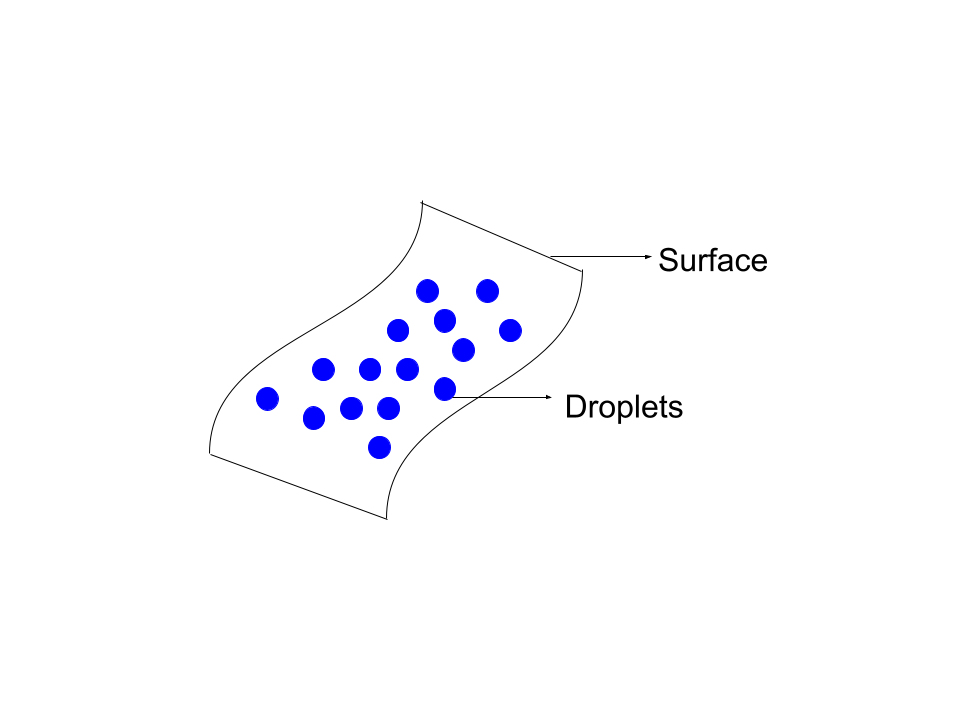}
    }
    \subfloat[]{
    \includegraphics[trim=5cm 5cm 0cm 6cm,width=0.5\textwidth]{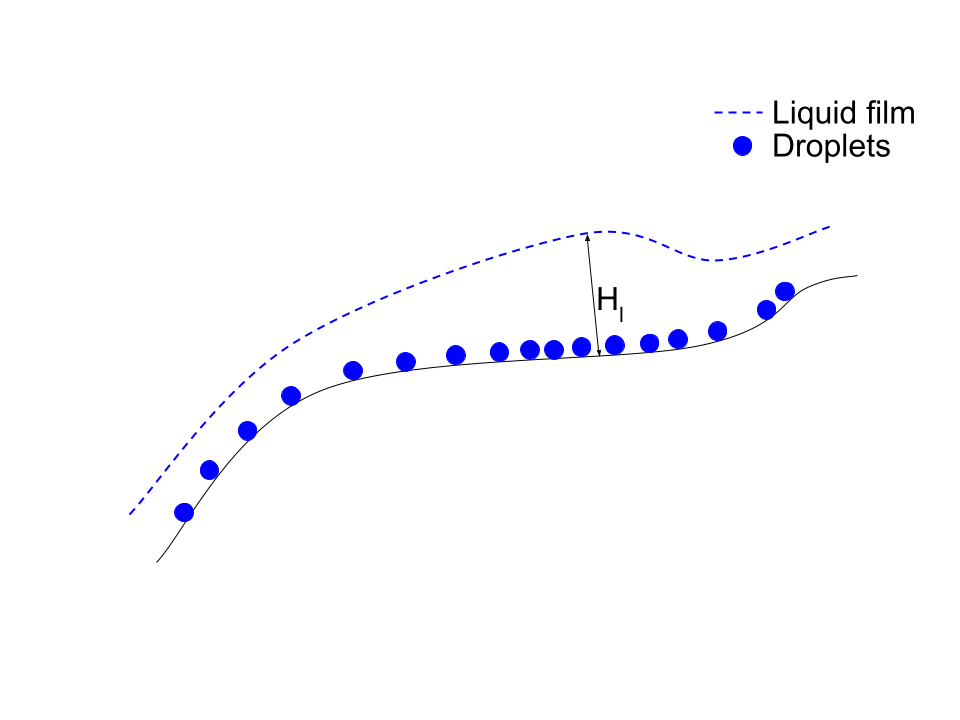}}
    \caption{\blue{Discrete Droplet Model (DDM): (a) A surface in $\mathbb{R}^3$ with an aggregation of droplets (b) Liquid film height determined from droplet locations on a surface.}}
    \label{fig:DDM}
\end{figure}
In the discrete droplet method (DDM) \cite{bharadwaj2022discrete}, a thin liquid film forming over a surface in $\mathbb{R}^3$ is discretized using droplets of prescribed diameters as shown in Fig.~\ref{fig:DDM}(a). The surface is given by a triangulated mesh with a set of consistently oriented surface normals. The DDM is a pseudo-2D model, in which the droplets move along the surface and the film height accounts for the third dimension. The aggregation of droplets represents the liquid thin-film as shown in Fig.~\ref{fig:DDM}(b) and is used to estimate the film height at a given location on the surface, as shown in the equation below. 
\begin{equation}
    H_{l,i} = \sum_{j \in S_i} W_{ij} V_j 
\end{equation}
Here, $H_{l,i}$ is the height of the liquid film at droplet $i$, $S_i$ is the neighborhood of droplet $i$ determined by proximity, $W_{ij}$ is the weight associated with droplet $j$ depending on its position with respect to droplet $i$, and $V_j$ is the volume of the droplet $j$. As shown in Fig.~\ref{fig:DDM}, in regions where there is more clustering of the droplets on the surface, the reconstructed height function, $H_l$, is higher. The derivatives of height and velocity are estimated using an SPH-like approach that uses Gaussian kernel functions. The neighbourhoods $S_i$ required for the derivative computations are obtained from search algorithms commonly used in meshfree methods\cite{dominguez2011neighbour,drumm2008finite}. \blue{The model solves the momentum equations at the droplet locations and accounts for the shear forces at the interface of the thin-film and the surface and the pressure gradients as a function of the gradients of the film-height. }For further details of the model, the work of Bharadwaj et al \cite{bharadwaj2022discrete} may be referred. 

\subsection{An overview of the new solidification model (DDM-S)}
In the proposed solidification model, DDM-S, we consider a liquid thin-film and a solidified region over a surface in $\mathbb{R}^3$. The basic components of the model are presented below.  
\begin{itemize}
    \item The discretization of the liquid thin-film is still done using droplets, just as in the case of the DDM. (See Sec.~\ref{sec:DDM})
    \item The liquid height $H_l$ is computed as in the case of the DDM. (See Sec.~\ref{sec:DDM})    
    \item In addition, we compute a solid height, $H_s$, as an attribute/property of a droplet. Consequently, a solidified region cannot be represented unless liquid droplets are present. Droplets are not located directly on the surface. Instead, they are at a distance $H_s = H_s(\vec{x}, t)$ away from the surface, as shown in Fig.~\ref{fig:solid_height}. (See Sec.~\ref{sec:solid height})
    \item For the evolution of the solid height $H_s$, an interface speed is evaluated in a 1-D subdomain, spanning between the surface and the edge of the liquid thin-film, by solving the Stefan problem. This interface speed governs the actual phase change process. (See Sec.~\ref{sec:s5}) 
    \item The droplet volume is then updated based on the change in liquid height. This step accounts for the volumetric change during solidification. (See Sec.~\ref{sec:s6})
    \item Unlike in the DDM, the liquid droplets move on the solidified region, along the solid-liquid interface, as shown in Fig.~\ref{fig:solid_height}. In DDM, droplets move along the surface, with the velocity tangential to the surface. Here, in DDM-S, droplets moving along the solid-liquid interface, with the velocity tangential to this interface. 
    Thus, the normals to the solid-liquid interface are computed at each droplet location and at each time step. (See Sec.~\ref{sec:s7})
    \item  As droplets move over the solid region, the solid height property, $H_s$, is updated using its derivatives in an advection equation. (See Sec.~\ref{sec:s1}) 
\end{itemize}
Each of these aspects of the model are expanded in detail in the subsections that follow.

\subsection{Solid height and solid-liquid interface}\label{sec:solid height}

\begin{figure}
    \centering
    \includegraphics[trim=0 2cm 1cm 1cm,width=0.8\textwidth]{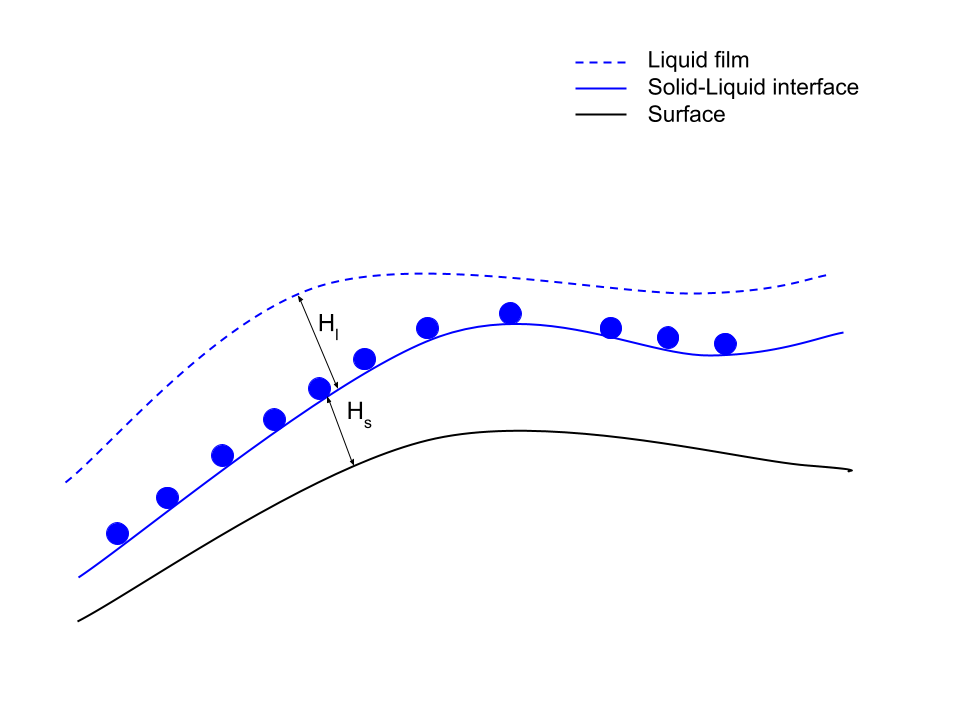}
    \caption{Modelling of solid height and solid-liquid interface: $H_l$ indicates the liquid height and $H_s$ indicates the solid height.}
    \label{fig:solid_height}
\end{figure}

    
The solid height at a droplet location, denoted as $H_s$, is stored as a property of the droplet.  The droplets are initialized with a solid height field at the beginning of every simulation. The solid-liquid interface, thus, is located at a distance $H_s$ away from the surface, along the surface normal, as shown Fig. \ref{fig:solid_height}. Unlike the liquid film height ($H_l$), the solid height ($H_s$) is not obtained by summing the Gaussian functions of the droplets in the neighbourhood.  Instead, it is treated as a field property which is initialized and evolved at the droplet locations. The evolution of $H_s$ occurs due to 
\begin{enumerate}
    \item freezing/melting, as governed by the Stefan problem (Secs.~\ref{sec:s5}, \ref{sec:s6}).
    \item movement of droplets over the stationary solidified region that is captured by the advection equation that accounts for the velocity of the droplet and the gradient of solid height at the droplet location (Sec.~\ref{sec:s1}).
\end{enumerate}

Note that since the solid region is stored only as a property at the droplet locations, without droplets, a solid region cannot be represented by this model. However, this does not imply that a liquid thin-film must always be present to model a solidified region. For instance, an aggregation of droplets with zero volume each would represent a liquid thin-film of zero height, and can store a non-zero solid height ($H_s$).
\subsection{Determination of interface speed: Stefan problem for phase-change}

\begin{figure}
    \centering
    \includegraphics[trim=0 3cm 1cm 6cm,width=\textwidth]{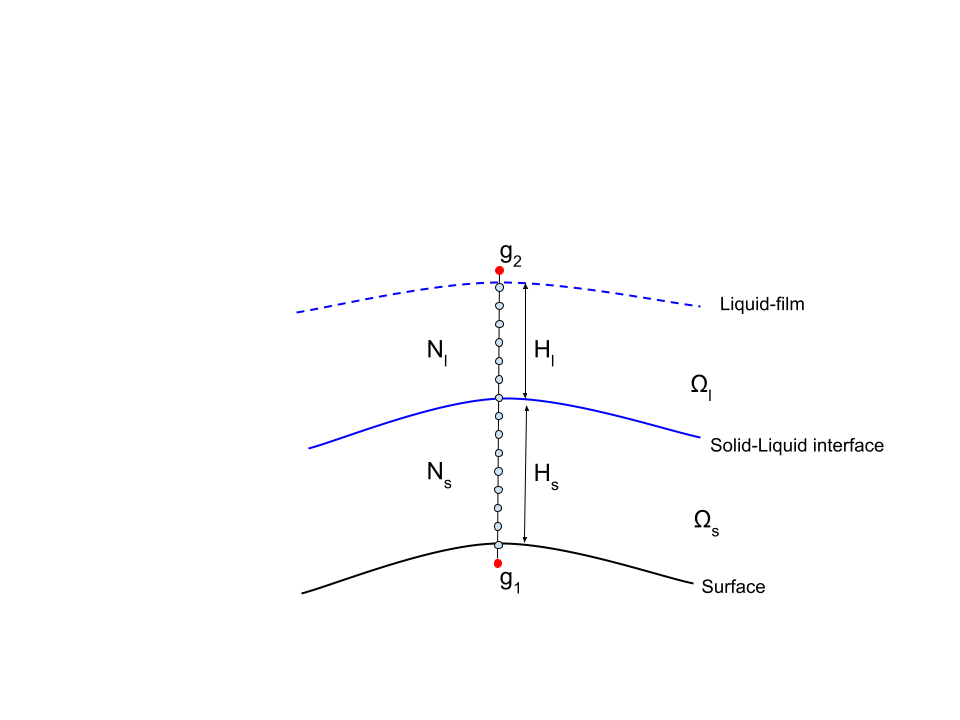}
    \caption{Evaluation of interface speed using a 1-D subdomain:  The red points are  ghost points used in the evaluation of energy equation on the liquid side ($\Omega_l$) and the solid side ($\Omega_s$)}
    \label{fig:1D_subdomain}
\end{figure}

In applications relating to freezing of liquid thin-films, the temperature gradients are typically much higher in the surface normal direction in comparison to the tangential directions, thus, resulting in solidification predominantly in the surface normal direction.  We, thus, adopt a pseudo-1-D approach that accounts for the surface normal gradients of temperature, to solve the phase-change problem at each droplet location. This is done by assuming a temperature profile at the droplet location, in a manner similar to assuming a velocity profile in a liquid thin-film required for viscous force calculations.\\
At a given droplet location, we consider a 1-D domain that lies between the surface and the edge of the liquid thin-film, as shown in Fig \ref{fig:1D_subdomain}. The solid-liquid interface is located at a distance $H_s$ from the surface. The energy equation is solved on either sides of this interface using a numerical discretization of $N_l$ points on the liquid side and $N_s$ points on the solid side, see Fig \ref{fig:1D_subdomain}.
\subsubsection{Stefan problem} \label{sec:s5}
The Stefan problem was first solved for a phase-change problem by J Stefan \cite{stefan1889einige}. We first derive the expression for the speed of the solid-liquid interface from Rankine--Hugoniot (R-H) jump conditions for conservation laws \cite{myers2020stefan}. Consider a conservation law of the form 
\begin{equation}
    \frac{\partial Q}{\partial t} + \nabla \cdot \vec{F} = 0
\end{equation}
 Here, $Q$ is the conserved quantity, and $\vec{F}$ is the flux. This equation is applicable in the two phases that are separated by a phase-change interface.
The speed of this interface, $v_I$, is given by 
\begin{equation}
    v_I = \frac{[F] \cdot \hat{n}}{[Q]}  
    \label{eq:RH}
\end{equation}
In the above equation, the pair of square brackets, $[\text{ }]$, denotes the jump in the enclosed quantity across the solid-liquid interface and $\hat{n}$ is the unit normal vector of the interface. 

In the present model, we use the R-H conditions to derive the solid-liquid interface speed assuming the flow is incompressible in each phase. Furthermore, the energy equation with only conduction is considered in each phase. The R-H condition is applied to both the mass and energy conservation equations. The continuity equation can be written as
\begin{equation}
    \frac{\partial \rho}{\partial t} + \nabla \cdot (\rho \vec{V}) = 0
\end{equation}
which is solved in both solid and liquid phases. Here, $\rho$ is the density of the phase and $\vec{V}$ is the velocity of the phase.
On applying the R-H condition, Eq.~\ref{eq:RH}, across the interface, the interface speed is estimated as
\begin{equation}
    v_I = \frac{[\rho \vec{V}] \cdot \hat{n}}{[\rho]} = \frac{(\rho_L \vec{V}_L - \rho_S \vec{V}_S)}{\rho_L - \rho_S} \cdot \hat{n}
\end{equation}
Here the subscripts $L$ and $S$ denote the liquid phase and the solid phase respectively.
In the present model, the velocity of the solid phase is zero since it is assumed to be stationary. Therefore, 
\begin{equation}
    v_I = \frac{\rho_L \vec{V}_L}{\rho_L - \rho_S} \cdot \hat{n} \implies \vec{V_L} \cdot \hat{n} = v_I \frac{\rho_L - \rho_S}{\rho_L}
    \label{eq:cont_step}
\end{equation}
The R-H condition is now considered for the energy equation, which can be written as
\begin{equation}
    \frac{\partial \rho E_t }{\partial t} + \nabla \cdot (\rho E_t \vec{V}  ) = -\nabla \cdot \vec{q}
    \label{Eq:energy}
\end{equation}
which is also solved in both solid and liquid phases.
Here, $E_t$ is the internal energy and $\vec{q}$ is the heat transfer due to conduction in the considered phase. Once again, applying the R-H condition, Eq.~\ref{eq:RH}, at the solid-liquid interface we arrive at
\begin{equation}
    v_I = \frac{[\rho E_t \vec{V} + \vec{q}] \cdot \hat{n}}{[\rho E_t]} = \frac{(\rho_L E_{t,L} \vec{V}_L + \vec{q}_L - \rho_S E_{t,S} \vec{V}_S - \vec{q}_S)}{\rho_L E_{t,L} - \rho_S E_{t,S}} \cdot \hat{n}
\end{equation}
Setting the solid phase velocity to zero and regrouping, 
\begin{equation}
v_I (\rho_L E_{t,L} - \rho_S E_{t,S}) =  \rho_L E_{t,L} (\vec{V}_L \cdot \hat{n}) + (\vec{q}_L - \vec{q}_S) \cdot \hat{n}
\end{equation}
Substituting  for $\vec{V_L} \cdot \hat{n}$ from Eq.~\ref{eq:cont_step} and regrouping, 
\begin{equation}
    v_I\left(\rho_L E_{t,L} - \rho_S E_{t,S} -  \rho_L E_{t,L}  \frac{\rho_L - \rho_S}{\rho_L} \right) =  (\vec{q}_L - \vec{q}_S) \cdot \hat{n}
\end{equation}
\begin{equation}
    v_I \rho_S (E_{t,L}-E_{t,S}) = (\vec{q}_L - \vec{q}_S) \cdot \hat{n}
\end{equation}
The interface is assumed to be at the melting temperature. Therefore, the difference in the energy $ (E_{t,L}-E_{t,S})$ at the interface, is only due to latent heat, $L$. Substituting for the heat fluxes using the Fourier's law,
\begin{equation}
    v_I = \frac{(\vec{q}_L - \vec{q}_S)}{\rho_S L} \cdot \hat{n} = \frac{1}{\rho_S L} \left(-\kappa_L \frac{\partial T}{\partial n}\Big|_L + \kappa_S \frac{\partial T}{\partial n}\Big|_S\right)
    \label{Eq:SC}
\end{equation}
Here $\kappa$ denotes the thermal conductivity. Using Eq.~\ref{Eq:SC}, the interface speed is computed from the jump in the temperature gradients across the interface. To evaluate the temperature gradients, we solve the energy equation, Eq.~\ref{Eq:energy}, in the two phases, as a 1-D heat equation at each droplet location. This is the case since the underlying flow is assumed to be incompressible. The solution to the 1-D heat equation is explained in Sec.~\ref{sec:1DHeatEquation}. 
\subsubsection{Temperature profile at droplet locations}\label{sec:temp_profile}

For each droplet, a temperature profile $T(n)$ needs to be reconstructed, as a function of the normal coordinate $n$ which denotes the distance from the surface in the normal direction. The temperature profile serves as an initial condition for the heat equation described in the previous subsection. At the solid-liquid interface, the temperature is equal to the melting temperature. At the solid -- surface interface and ambient-medium -- liquid-film interface, the temperature satisfies a convective heat transfer condition. Additionally, the height averaged temperature distribution in the liquid thin-film should equal the droplet temperature. These conditions are expressed mathematically as 
\begin{equation}
    T(n=H_s) = T_m
    \label{cond1}
\end{equation}
\begin{equation}
    \frac{\partial T}{\partial n} \Big|_{n=0} = \frac{h_s}{\kappa_s}(T(0)-T_s)
    \label{cond2}
\end{equation}
\begin{equation}
    \frac{\partial T}{\partial n} \Big|_{n=H_s+H_l} = -\frac{h_a}{\kappa_l}(T(H_s+H_l)-T_a)
    \label{cond3}
\end{equation}
\begin{equation}
    \frac{1}{H_l}\int_{H_s}^{H_l} T(n) dn = T_{avg} 
        \label{cond4}
\end{equation}
In the above equations, $T_m$ is the melting/freezing temperature, $T_s$ is the surface temperature, $T_{avg}$ is the average droplet temperature and $h_s$ and $h_a$ are the heat transfer coefficients at the surface-solid interface and ambient medium-liquid interface respectively. \blue{It is noted that spatial and temporal variations in $T_s$ can be accommodated through Eq.~\ref{cond2}.} In the liquid region, there are three conditions that are to be satisfied by the temperature profile -- temperature at the solid-liquid interface (Eq.~\ref{cond1}), heat transfer between the liquid-film and the ambient medium (Eq.~\ref{cond3}) and the average temperature of the droplet (Eq.~\ref{cond4}). Thus, 
the temperature profile in the liquid region is assumed to follow a quadratic curve of the form 
\begin{equation}
    T(n) = a_l n^2 + b_l n + c_l \text{ , } H_s< n \le H_l 
    \label{eq:tprofile1}
\end{equation}
The constants $a_l$, $b_l$ and $c_l$ can be evaluated from conditions of Eqns.~\ref{cond1},\ref{cond3} and \ref{cond4}. \\
The temperature profile in the solid region is assumed to follow a linear curve, owing to the presence of two conditions -- the temperature at the solid-liquid interface (Eq.~\ref{cond1}) and the heat transfer between the solid region and the underlying surface (Eq.~\ref{cond2}).
\begin{equation}
        T(n) = a_s n+ b_s \text{ , } 0 \le n \le H_s
        \label{eq:tprofile2}
\end{equation}
The constants $a_s$ and $b_s$ can be evaluated from conditions of Eqns.~\ref{cond1} and \ref{cond2}. The temperature profile is, thus, continuous at the solid-liquid interface with its value equal to the melting temperature, but the normal derivative of temperature could have a discontinuity. It is this discontinuity in the derivative that causes the movement of the solid-liquid interface, as seen in Eq.~\ref{Eq:SC}. The derivation of the constants involved in the temperature profile $(a_l,b_l,c_l)$ and $(a_s,b_s)$ are elaborated in ~\ref{AppA}. 

 It is reiterated that the temperature profile obtained as described here, serves as an initial condition. As the simulation progresses, the change in the temperature profile is given by the heat equation described in Sec.~\ref{sec:1DHeatEquation}.

\subsubsection{Numerical solution of the energy equation in a 1-D sub-domain with an evolving interface}\label{sec:1DHeatEquation}

At each droplet location, the temperature gradients on the liquid side and the solid side are needed, as shown in Eq.~\ref{Eq:SC}. The heat equation is, thus, solved at each droplet location in a 1-D subdomain discretized by $N$ points, see Fig \ref{fig:1D_subdomain}. 
\begin{equation}
    \frac{\partial T}{\partial t} = \frac{\kappa}{\rho c} \frac{\partial ^2 T}{\partial n^2} = \alpha  \frac{\partial ^2 T}{\partial n^2}
\label{eq:heat_equation}
\end{equation}
Here, $c$ denotes specific heat and $\alpha$ denotes the thermal diffusivity.
The initial condition at each point is provided by the analytical form described in Sec.\ref{sec:temp_profile}. The physical extent of the 1-D sub-domain begins at the surface and ends at the edge of the liquid film. There are two ghost points on either side, as shown by the red points in Fig \ref{fig:1D_subdomain}, to facilitate the imposition of boundary conditions. The solid-liquid interface is at the melting temperature and is used as a boundary condition for the heat equation on both the solid and liquid phases. 
The temperatures at the two ghost points ($g1$ and $g2$) are determined based on the convective heat transfer boundary condition at the surface and edge of the liquid film, as shown below:
\begin{equation}
    Q_s = \kappa \frac{\partial T}{\partial n}\Big|_s = h_s (T_{g1}-T_s)
\end{equation}
\begin{equation}
        Q_a = \kappa \frac{\partial T}{\partial n}\Big|_a = -h_a (T_{g2}-T_a)
\end{equation}
Here, $h_s$ and $h_a$ are the convective heat transfer coefficients at the surface and at the interface with the ambient fluid respectively. $T_s$ and $T_a$ are the surface temperature and the temperature of the ambient fluid respectively. $Q_s$ and $Q_a$ denote the heat transfer rate due to the convective boundary condition at the respective boundary. \\
Eq.~\ref{eq:heat_equation} is solved using a finite difference method. 
\begin{equation}
    T_k^{m+1} (1+2\lambda) - T_{k+1}^{m+1} \lambda - T_{k-1}^{m+1} \lambda= T_k^m
        \label{eq:discrete_eqn}
\end{equation}
where $\lambda = \frac{\alpha \Delta t}{\Delta n^2} $, and $T_k^m$ denotes the temperature $T$ at point $k$ of the 1-D sub-domain at time level $m$.. Near the boundaries and in the vicinity of the solid-liquid interface, the above equation is modified to impose the boundary conditions. For instance, at the first and last points of the 1-D sub-domain, the discretization is shown below. 
\begin{equation}
    T_1^{m+1} (1+2\lambda) - T_{2}^{m+1} \lambda = T_1^m + T_{g1} \lambda
\end{equation}
\begin{equation}
    T_N^{m+1} (1+2\lambda) - T_{N-1}^{m+1} \lambda = T_N^m + T_{g2} \lambda
\end{equation}
For the point that lies on the solid-liquid interface, the melting temperature is forced.
\begin{equation}
    T_{N_s} = T(n=H_s) = T_m 
\end{equation}
These equations lead to a system of linear equations with a tridiagonal matrix of the form\\

\begin{equation}
    \begin{bmatrix}
        1+2\lambda & -\lambda   &  0       & . & . &. & .& .& .& .&. \\
        -\lambda  & 1+2\lambda & -\lambda & 0 & . &. &. & .& .& .&. \\
        .  &  &  & & & & & & & & \\
        .  &  &  & & & & & & & & \\
        .  &  &  & & & & & & & & \\
        .  &  &  .&. &0 & 1 & 0&. & .& & \\
                .  &  &  & & & & & & & & \\
        .  &  &  & & & & & & & & \\
        .  &  &  & & & & & & & & \\
                . & . &  .      & . & . &. & .& .& -\lambda& 1+2\lambda & -\lambda  \\
        .  & . & . & . & . &. &. & .& 0& -\lambda &1+2\lambda \\
    \end{bmatrix}
            \begin{bmatrix}
        T_1^{m+1}   \\
        T_2^{m+1}  \\
        .  \\
        .   \\
        .  \\
        T_{N_s}  \\
                .  \\
        .   \\
        .   \\
        T_{N-1}^{m+1}\\
        T_{N}^{m+1}    \\
    \end{bmatrix}
    =
        \begin{bmatrix}
        T_1^m + T_{g1} \lambda  \\
        T_2^m  \\
        .  \\
        .   \\
        .  \\
        T_m  \\
                .  \\
        .   \\
        .   \\
        T_{N-1}^m\\
        T_{N}^m + T_{g2} \lambda    \\
    \end{bmatrix}
\end{equation}
The above system of equations is solved using the LSQR algorithm. The number of points in the 1-D sub-domain may be chosen such that it sufficiently resolves the temperature gradients. \blue{The use of too few points may adversely affect the solution.} In the test cases presented in this paper, the total number of points in the 1-D subdomain is 20 with 10 points discretising the solid region and 10 points discretising the liquid region. For these test cases, using a larger number of points does not change the solution.

\blue{By solving for the temperature field from the heat equation, as described here, this model accommodates for unsteadiness in the temperature distribution and relaxes assumptions imposed by Myers' model \cite{myers2001extension} about the thermal problem being quasi-steady and the temperature profile being linear.}

\subsubsection{Temperature interpolation due to change in droplet neighborhood}
 \begin{figure}
    \centering
    \includegraphics[trim=0 3cm 1cm 6cm,width=\textwidth]{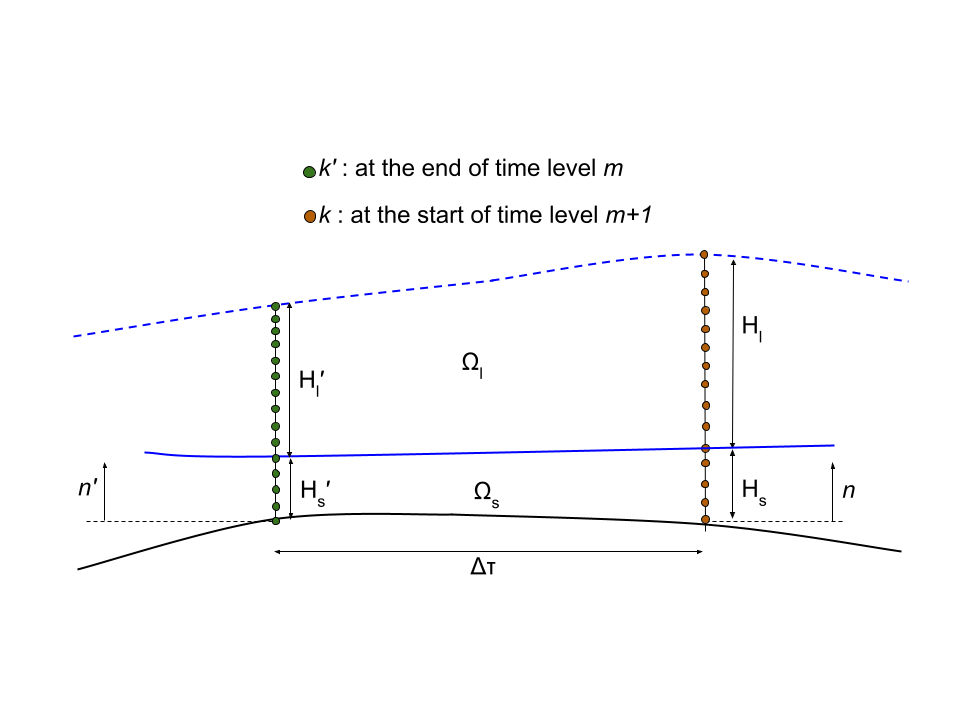}
    \caption{A schematic representing a scenario of temperature projection when a droplet moves or neighbourhood changes between time-steps. The primed quantities are before and non-primed are after the droplet movement/change in neighbourhood has occurred. }
    \label{fig:temperature projection}
\end{figure}
In Eq.~\ref{eq:discrete_eqn}, $T^m$ denotes the temperature of the droplet at its location at the beginning of time level $m+1$. However, this temperature is not directly available since the previously stored temperature profile corresponds to the droplet's previous location and neighborhood, before the Lagrangian motion. Therefore, the temperature profile at the end of time level $m$ needs to be projected to the current droplet locations, to accommodate its new location and new neighborhood before solving Eq.~\ref{eq:discrete_eqn}. We denote the known temperature profile before the droplet movement as $T^{k'}$ and the projected temperature profile as $T^k$ (corresponds to $T^m$ of Eq.~\ref{eq:discrete_eqn}). \\
This process is illustrated in Fig.~\ref{fig:temperature projection}. We first consider the solid region. Let the normal coordinate before projection be denoted as $n'$ and after the projection be denoted as $n$ . The projected temperature profile of the 1-D sub-domain is given as
\begin{equation}
    T^{k}(n) = T^{k'}(n') + \frac{\partial T}{\partial \tau}(n')\Big|_{k'} \Delta \tau
    \label{eq:tempvar1}
\end{equation}
where $\frac{\partial T}{\partial \tau}$ is the variation of temperature in the tangential direction and $\Delta \tau$ is the distance moved by the droplet on the tangent plane. We assume a linear variation of $\frac{\partial T}{\partial \tau}$ w.r.t $n'$. The gradient at the wall $\frac{\partial T}{\partial \tau}\Big|_{n'=0} = q_1$ and at the solid-liquid interface $\frac{\partial T}{\partial \tau} \Big|_{n'=H'_s} = 0$, serve as boundary conditions for the linear profile. This gives
\begin{equation}
    \frac{\partial T}{\partial \tau} (n') = q_1 \Big(1-\frac{n'}{H'_s}\Big) 
\end{equation}
Substituting this in Eq.~\ref{eq:tempvar1}, 
\begin{equation}
    T^{k}(n) = T^{k'}(n') + q_1 \Big(1-\frac{n'}{H'_s}\Big)  \Delta \tau
\end{equation}
Secondly, we consider the liquid region. We proceed along similar lines as in the solid region, 
\begin{equation}
    T^{k}(n) = T^{k'}(n') + \frac{\partial T}{\partial \tau}(n')\Big|_{k'} \Delta \tau
    \label{eq:tempvar2}
\end{equation}
In this case, however, the normal coordinate is measured starting from the solid-liquid interface. The boundary conditions for the linear profile of $\frac{\partial T}{\partial \tau}$ is $\frac{\partial T}{\partial \tau} \Big|_{n'=0} = 0$ and $\frac{\partial T}{\partial \tau} \Big|_{n'=H'_l} = \gamma$. This gives, 
\begin{equation}
    \frac{\partial T}{\partial \tau} (n') = \frac{\gamma n'}{H'_l}
\end{equation}
Substituting this in Eq.~\ref{eq:tempvar2}, we get 
\begin{equation}
    T^{k}(n) = T^{k'}(n') + \frac{\gamma n'}{H'_l} \Delta \tau
    \label{eq:liquidtemperture}
\end{equation}
Usually, the tangential derivative, $\gamma$ at $n'=H'_l$ is unknown. The normal derivative, on the other hand, can be estimated from the convective heat transfer occurring between the liquid-film and the surroundings. Assuming this to be equal before and after projection,
\begin{equation}
    \frac{\partial T}{\partial n}\Big|_{n=H_l} =  \frac{\partial T}{\partial n}\Big|_{n'=H'_l} = q_2
    \label{eq:convBC}
\end{equation}
Differentiating Eq.~\ref{eq:liquidtemperture} w.r.t $n$, we get
\begin{equation}
    \frac{\partial}{\partial n} T^{k}(n) = \frac{\partial}{\partial n'} T^{k}(n) \frac{\partial n'}{\partial n} = \Big[\frac{\partial}{\partial n'} T^{k}(n') + \frac{\gamma}{H'_l}\Delta \tau \Big] \frac{H'_l}{H_l}
\end{equation}
In the above equation, $\frac{\partial n'}{\partial n}$ denotes the scaling of the normal coordinate which is given by the ratio of the liquid heights $\frac{H'_l}{H_l}$.
Now, setting $n=H_l$ and substituting from Eq.~\ref{eq:convBC}, we get
\begin{equation}
    q_2 = \Big[q_2 + \frac{\gamma}{H'_l}\Delta \tau\Big] \frac{H'_l}{H_l}
\end{equation}
\begin{equation}
\gamma = q_2\Big(1-\frac{H'_l}{H_l}\Big) \frac{H'_l}{\Delta \tau}
\end{equation}
Substituting for $\gamma$ in Eq.~\ref{eq:liquidtemperture}, we get the below equation that is used to project the temperature in the liquid region of the 1-D sub-domain. 
\begin{equation}
      T^{k}(n) = T^{k'}(n') + q_2 n'\Big(1-\frac{H'_l}{H_l}\Big)
      \label{eq:liquid_projected_temperature}
\end{equation}
\subsection{Updating heights, droplet repositioning and updating droplet volume} \label{sec:s6}
The movement of the interface, given by the Stefan problem (see Sec.~\ref{sec:s5}), corresponds to a change of the heights of the solid and liquid parts of the thin film, as shown below. 
\begin{equation}
    v_I = \frac{\partial H_s}{\partial t} = -\beta \frac{\partial H_l}{\partial t}
\end{equation}
In this equation $\beta$, which is the ratio of density of the solid to that of the liquid, accounts for volumetric expansion/contraction.  \\
 When the solid height changes based on the Stefan condition, the droplet needs to be re-positioned to the updated solid-liquid interface by moving it by the appropriate distance. 
 \begin{equation}
    \Delta \Vec{x} = \Delta H_s \vec{n}_s
\end{equation}
Here, $\vec{n}_s$ is the unit normal to the surface.
The volume of the droplet also needs to be updated depending on whether $H_s$ increases or decreases. For instance, if $H_s$ increases, the height of the liquid-film $H_l$ will reduce owing to mass conservation. This reduction in height is accounted for by a reduction in droplet volume. To derive an expression for the updated volume, a solid fraction is defined at the droplet location as 
\begin{equation}
    f = \frac{H_s}{H_s+H_l}
    \label{eq:solid_fraction}
\end{equation}
The volume of the droplet relates to the solid fraction as 
\begin{equation}
    V = (1-f) V_0
\end{equation}
where $V_0$ corresponds to the volume of the droplet when no solid is present (i.e. $f=0$). Therefore, 
\begin{equation}
    \Delta V = -V_0 \Delta f 
\end{equation}
The change in solid fraction can be estimated by using the old and new heights in Eq.~\ref{eq:solid_fraction}. 

\subsection{Updating solid normal}\label{sec:s7}
\begin{figure}
    \centering
    \includegraphics[trim=0 3cm 1cm 6cm,width=0.75\textwidth]{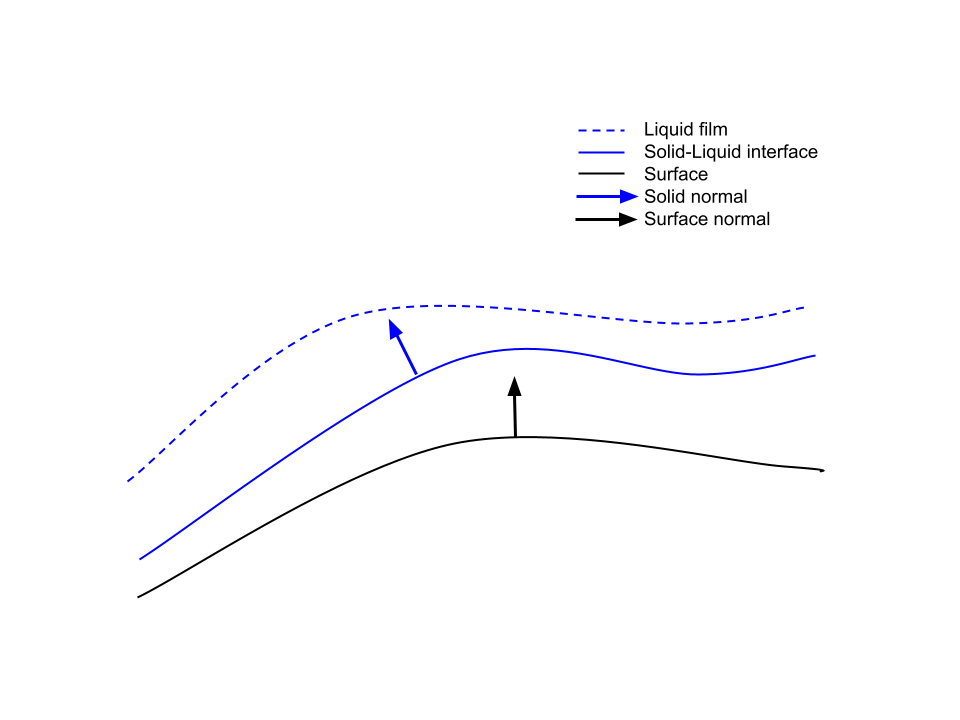}
    \caption{ Solid normal vs Surface normal: While DDM requires the surface normal for droplet movement, the DDM-S requires the solid normal.}
    \label{fig:solidvssurfacenormal}
\end{figure}

For the movement of the liquid thin-film over the solidified region, the surface normals of the conventional DDM need to be revisited. The normal to the solidified region, associated with a given droplet is referred to as the solid normal. The distinction between the solid and surface normal is illustrated in Fig.\ref{fig:solidvssurfacenormal}. \green{ While the surface normal is used for the solution of the heat equation in the 1-D subdomain, the solid normal is used for the movement of droplets over the solidified region.}
The solid normal is calculated based on the position of droplets in the neighbourhood of the given droplet. This procedure is elaborated in \cite{saucedo2019three}. For a given droplet $i$, the quantity $J$ shown below, needs to be minimized. $J$ denotes sum of the projections of the solid normal onto the solidified region. The solidified region is represented by the droplets in the neighbourhood of droplet $i$. When this sum, $J$, is minimized, the resulting vector $\vec{n}_i$ is the required solid normal. 
\begin{equation}
    J = \sum_{j \in S_i} W_{ij} (\vec{n}_i \cdot \vec{r}_{ij})^2
\end{equation}
Here, $\vec{n}_i$ is the unit solid normal at the droplet $i$ and $\vec{r}_{ij}$ is the vector joining droplets $i$ and $j$. This can be recast as 
\begin{equation}
    J = \Vec{n}^T C \vec{n}
\end{equation}
where the matrix $C$ is a $3 \times 3$ matrix is given by
\begin{equation}
\begin{bmatrix}
\sum_{j \in S_i} W_{ij} \Delta x^2 & \sum_{j \in S_i} W_{ij} \Delta x \Delta y & \sum_{j \in S_i} W_{ij} \Delta x \Delta z \\
\sum_{j \in S_i} W_{ij} \Delta x W_{ij} \Delta y & \sum_{j \in S_i} W_{ij} \Delta y^2 & \sum_{j \in S_i} W_{ij} \Delta y \Delta z \\
\sum_{j \in S_i} W_{ij} \Delta x W_{ij} \Delta z & \sum_{j \in S_i} W_{ij} \Delta z \Delta y & \sum_{j \in S_i} W_{ij} \Delta z^2 \\
\end{bmatrix}
\end{equation}
such that $\vec{r}_{ij} = \Delta x \text{ }\hat{i} + \Delta y\text{ } \hat{j} + \Delta z \text{ }\hat{k}$. A Gaussian function is used to evaluate the weight, $W_{ij}$, between droplets $i$ and $j$.
This minimization leads to an Eigen value problem. 
\begin{equation}
    {C} \vec{n} = \lambda \vec{n} 
\end{equation}
 
The normalized Eigen vector corresponding to the smallest Eigen value of the matrix $C$, is the required solid normal. 
\subsection{Advection of solid height} \label{sec:s1}
The solid height, as mentioned earlier, is treated as a property of the liquid droplet. Since the solidified region is stationary and the droplets move, the solid height cannot be advected along with the droplet with no change. Instead, the solid height needs to be revised as 
\begin{equation}
    \frac{\partial H_s}{\partial t}  -\vec{V}_{\text{drop}} \cdot \nabla H_s = 0
    \label{eq:solid_advection}
\end{equation}
Note the negative sign preceding the convective term. In other words, the solid region moves at a velocity $-\vec{V}_{\text{drop}}$ w.r.t the droplets.
The derivative of the solid height is given by an SPH-based formulation as
\begin{equation}
    \nabla H_s = -\sum_{j \in S_i} \nabla W_{ij} H_j \Delta A_j
\end{equation}
From the consistency condition $\sum_{j \in S_i} \nabla W_{ij}\Delta A_j = 0$, the above equation may be rewritten as
\begin{equation}
    \nabla H_s = -\sum_{j \in S_i} \nabla W_{ij} (H_j+H_i) \Delta A_j = -2\sum_{j \in S_i} \nabla W_{ij} \frac{(H_j+H_i)}{2} \Delta A
\end{equation}
$(H_i + H_j)/2$ can be interpreted as an average, $H_{ij}$, at the midpoint of the line joining droplets $i$ and $j$. 
\begin{equation}
    \nabla H_s = -2\sum_{j \in S_i} \nabla W_{ij} H_{ij} \Delta A
\end{equation}
Upwinding can now be performed on the term $H_{ij}$, to render the marching of the equation stable. 
\begin{align}
    H_{ij} = H_i \text{, if } (\vec{x}_i - \vec{x}_j) \cdot \vec{V}_{\text{drop}} > 0 \\
     =   H_j \text{, if } (\vec{x}_i - \vec{x}_j) \cdot \vec{V}_{\text{drop}} < 0
\end{align}
\blue{The evolution of solid height using the method described above makes it susceptible to dispersive and dissipative numerical errors. While smooth solid profiles can be satisfactorily captured, for profiles that vary sharply these errors need to be tackled by using sufficiently small temporal and spatial resolutions or higher-order discretizations.}
\subsection{Surface point and position correction} \label{sec:s2}
For each droplet moving on the solid, a corresponding point on the surface is stored. This is the surface point that lies closest to the droplet such that the line joining the point and the droplet is normal to the surface.
When a droplet moves over the solidified region, the solid height at the new location, $H_{s}$, may not necessarily match with the height obtained by the advection step, $H_{s}^{\text{adv}}$, described in Sec.~\ref{sec:s1}. \blue{ To ensure that the actual height , $H_{s}$, matches the height from the advection equation, $H_{s}^{\text{adv}}$, a correction is introduced in the position as shown below, where $\hat{n}_s$ is the normal to the surface. }
\begin{equation}
    \Delta \vec{x}_c = \hat{n}_s(H_{s}^{\text{adv}}-H_{s}) 
\end{equation}
\subsection{Impinging droplets} \label{sec:s3}
The present model incorporates handling of impinging droplets, which is an important aspect in  applications like aircraft or wind-turbine icing. When a liquid droplet moving in the bulk-flow meets the surface, it becomes a part of the thin-film model (DDM). At the location of this droplet, the solidified height is initialized by performing an interpolation from neighbouring droplets and an initial temperature profile is prescribed. \\
The solid height at the new droplet location is interpolated as
\begin{equation}
    H_{s,i}  = \sum_{j \in S_{i},j \neq i} W_{ij}H_{s,j}
\end{equation}
For the temperature profile at the new droplet location, the following boundary conditions are used to determine a quadratic temperature profile.
\begin{equation}
    T (n = H_s) = T_m 
\end{equation}
\begin{equation}
    \frac{\partial T}{\partial n} \Big|_{n=H_s+H_l} = -h_a[T(n=H_s+H_l)-T_a]
\end{equation}
\begin{equation}
    \frac{1}{H_l}\int_{H_s}^{H_l} T(n) dn = T_{avg} 
\end{equation}
 
Here, $T_m$ is the melting temperature, $T_a$ is the ambient temperature and $h_a$ is the convective heat transfer coefficient at the edge of the liquid film. 

\subsection{Summary of algorithm}
The entire DDM-S model is implemented in an in-house solver, MESHFREE\footnote{www.meshfree.eu}. A summary of steps of the algorithm is given below, with links to the appropriate sections above. 

\begin{algorithm}
    \caption{Discrete Droplet Method with Solidification} \label{alg:DDM-S}
    \begin{algorithmic}[1]
        \State Initialize solid height $H_s$ at the droplet locations. 
        \State Move the droplets from the surface to the solid-liquid interface 
        \While{Time-stepping loop}
            \State Lagrangian movement of droplets \cite{suchde2018point}
            \State Recomputing surface points - Sec.~\ref{sec:s2}
            \State Advection of solid height - Sec.~\ref{sec:s1}
            \State Position correction - Sec.~\ref{sec:s2}
            \State Impinging droplet treatment - Sec.~\ref{sec:s3}
            \State Stefan problem -  Evaluation of interface speed -  Sec.~\ref{sec:s5}
            \State Update heights, droplet volume and position - Sec.~\ref{sec:s6}
            \State Update solid normals - Sec.~\ref{sec:s7}
            \State Compute new \blue{droplet} velocity using DDM \cite{bharadwaj2022discrete}
        \EndWhile
    \end{algorithmic}
\end{algorithm}

\section{Results}
In this section, we present the results of the proposed model in stages of increasing complexity. The test cases are tabulated in Table.~\ref{tab:cases}. While test cases 1-3 verify individual parts of the model, the last two test cases involving a circular conductor and an aircraft wing incorporate all aspects of the model. In all the test cases, the ratio of densities of the solid and liquid, $\beta$, is set to unity implying no volumetric change. A validation of the underlying DDM model (without solification), can be found in our earlier work \cite{bharadwaj2022discrete}. 
\begin{table}
\centering
\caption{List of test cases}
\label{tab:cases}
\resizebox{\textwidth}{!}{
\begin{tabular}{ll}
\hline \hline
Test case   & Remarks \\ \hline
1: Thin-film of constant velocity sliding over a solid layer of varying height & Verifies advection of solid height \\
   2: Sliding thin-film of constant height over a solid layer of varying height & Verifies DDM solution in the presence of a solid region\\
   3: Solidification of a stationary thin film & Verifies the Stefan condition-based phase-change model \\
   4: Solidification on a circular conductor with droplet impingement & Validates the complete model\\
   5: Solidification on an aircraft wing with droplet impingement & Validates the complete model\\
   \hline \hline
\end{tabular}
}
\end{table}
\subsection{Test Case 1: Thin-film of constant velocity sliding over a solid layer of varying height  }
\begin{figure}
    \centering
    \subfloat[A sketch of the solid profile given by Eq.~\ref{eq:TC1}, over which  droplets slide at a constant velocity.]{
    \includegraphics[trim=0cm 6cm 0cm 10cm, width=0.5\textwidth]{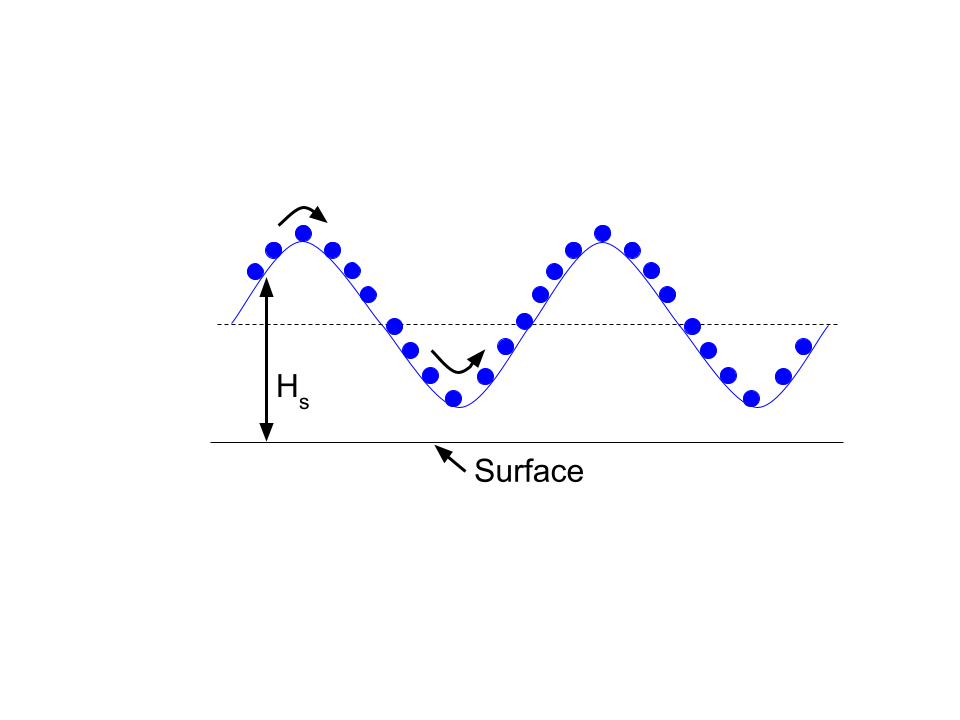}
    }
    \quad
    \subfloat[The numerical domain of computation: a flat plate initialized with a wavy solidified region.]{
    \includegraphics[trim=0cm 6cm 0cm 8cm, width=0.4\textwidth]{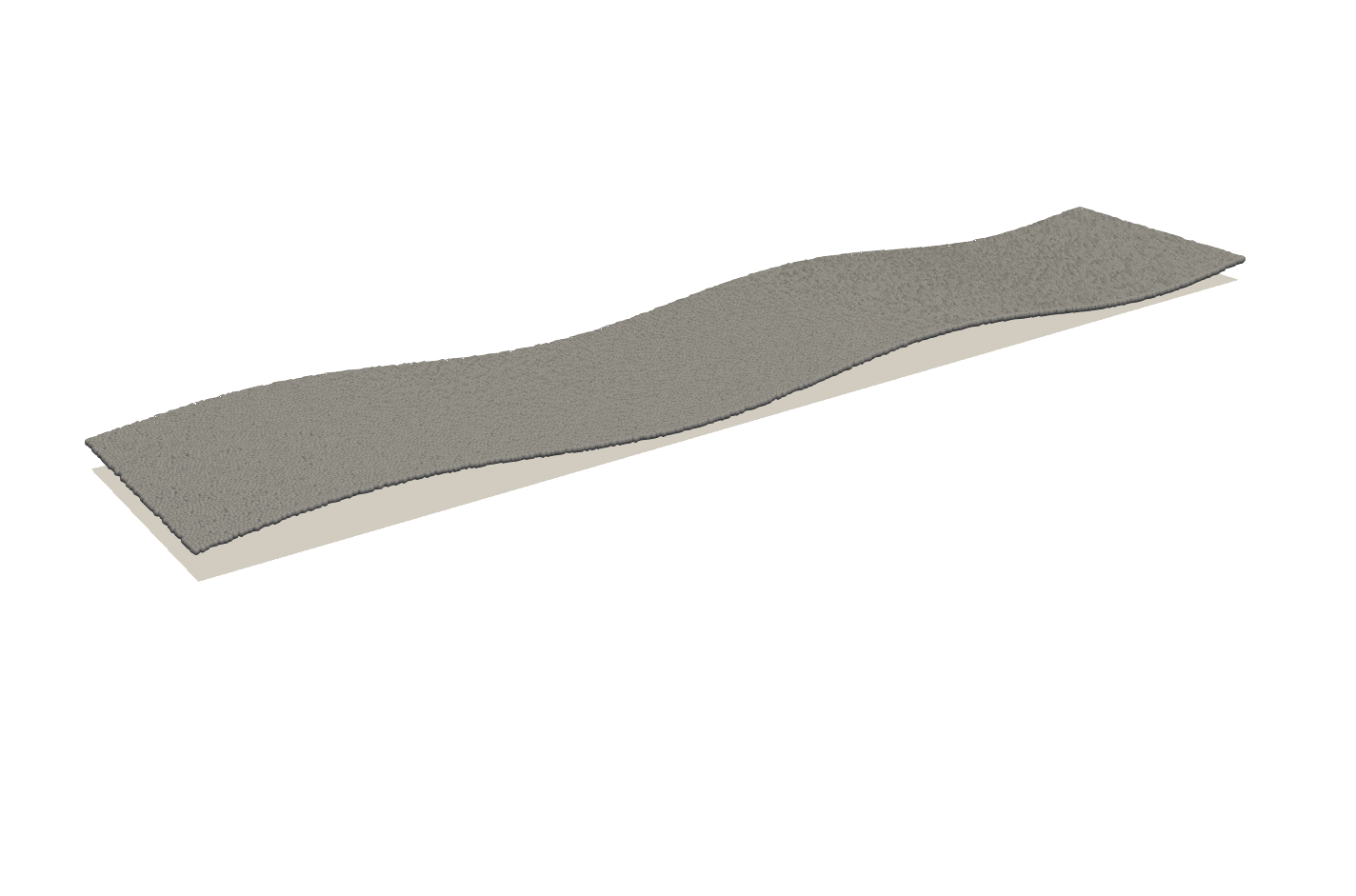}}
    \caption{Test Case 1: Domain of computation. }
    \label{fig:TC1_domain}
\end{figure}

\begin{figure}
    \centering
    \subfloat[t = 0s]{
    \includegraphics[width=0.5\textwidth]{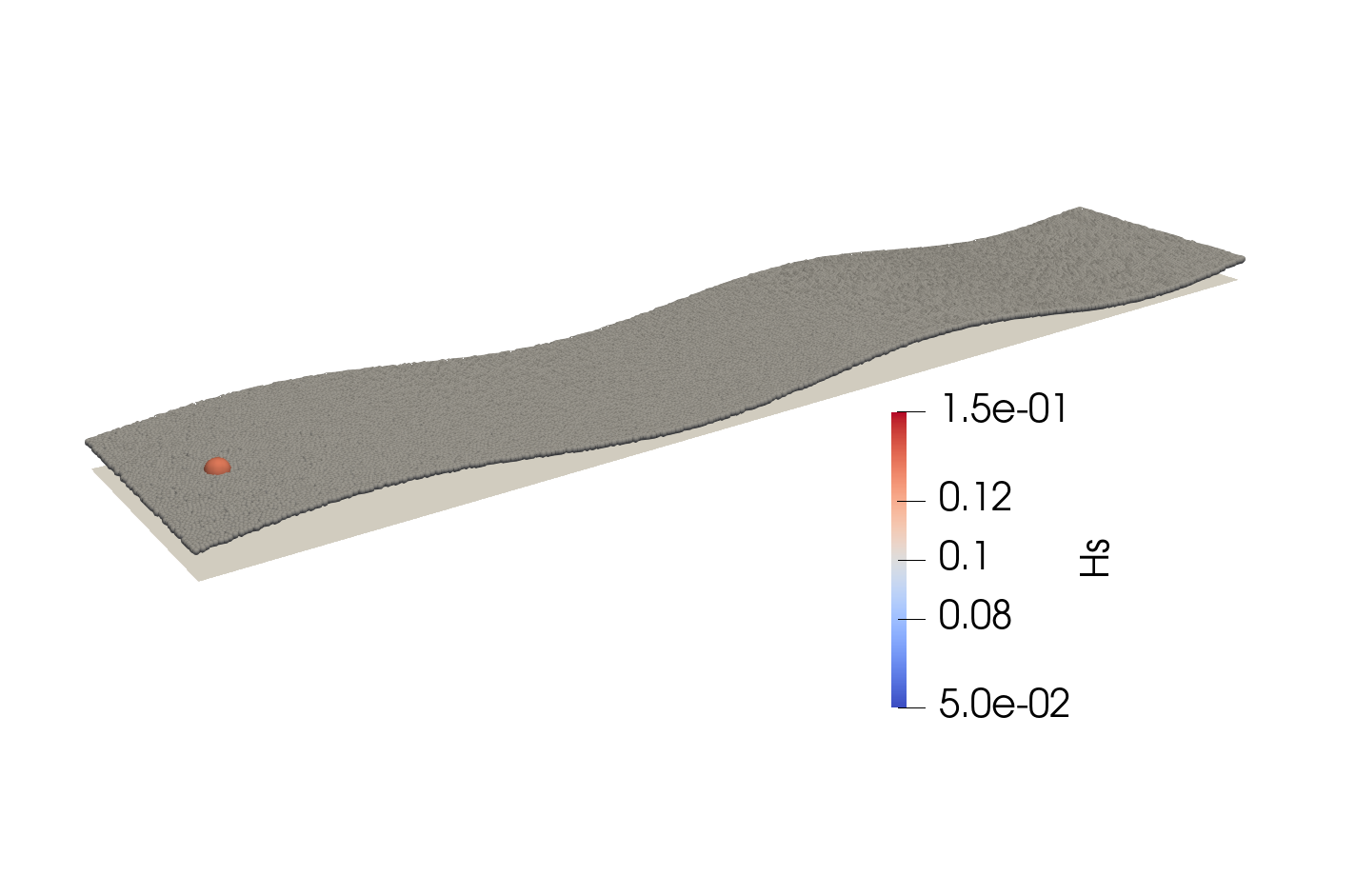}}
    \subfloat[t = 0.03s]{
    \includegraphics[width=0.5\textwidth]{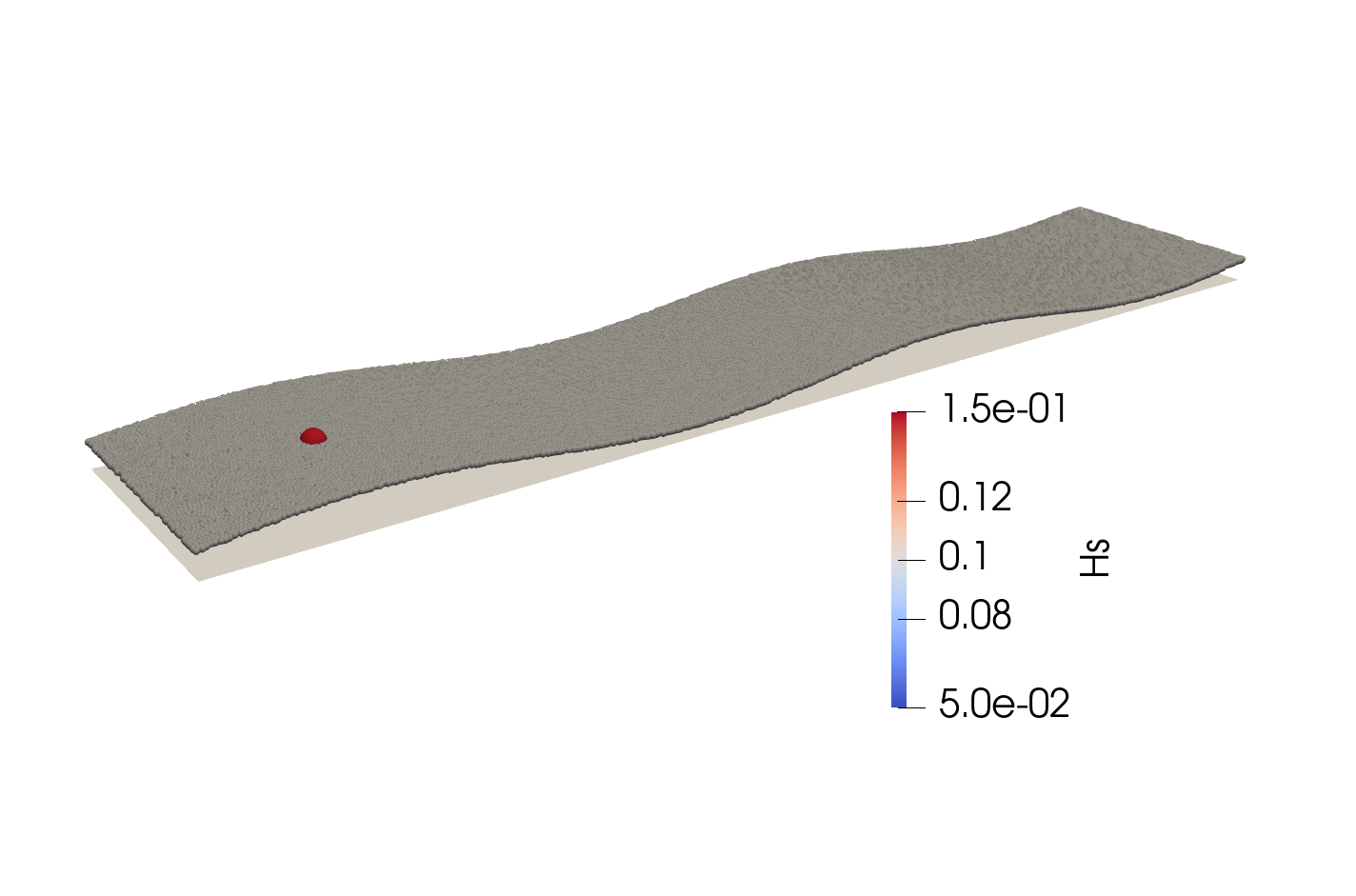}
    }\\
    \subfloat[t = 0.12s]{
    \includegraphics[width=0.5\textwidth]{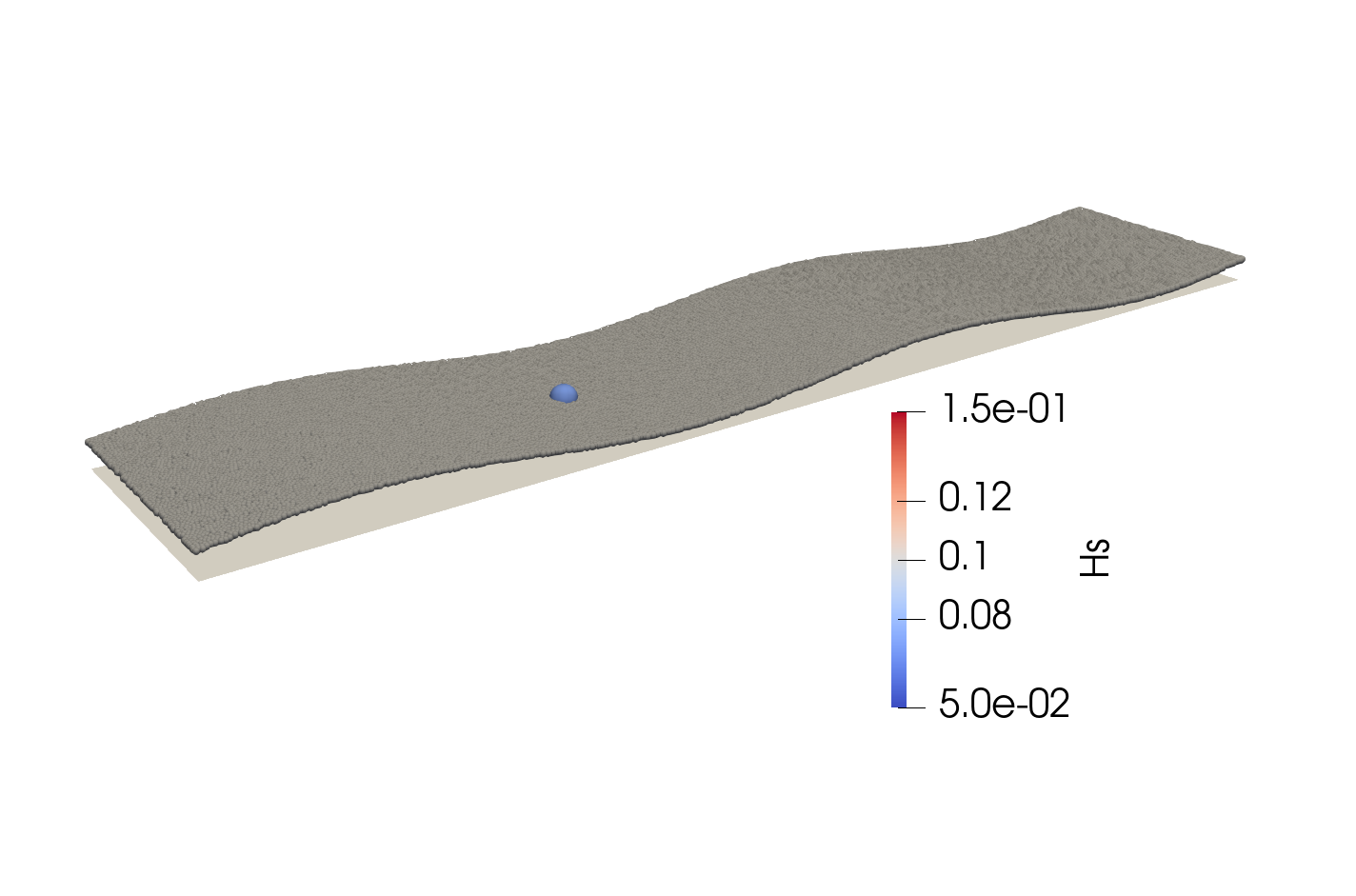}}
    \subfloat[t = 0.15s]{
    \includegraphics[width=0.5\textwidth]{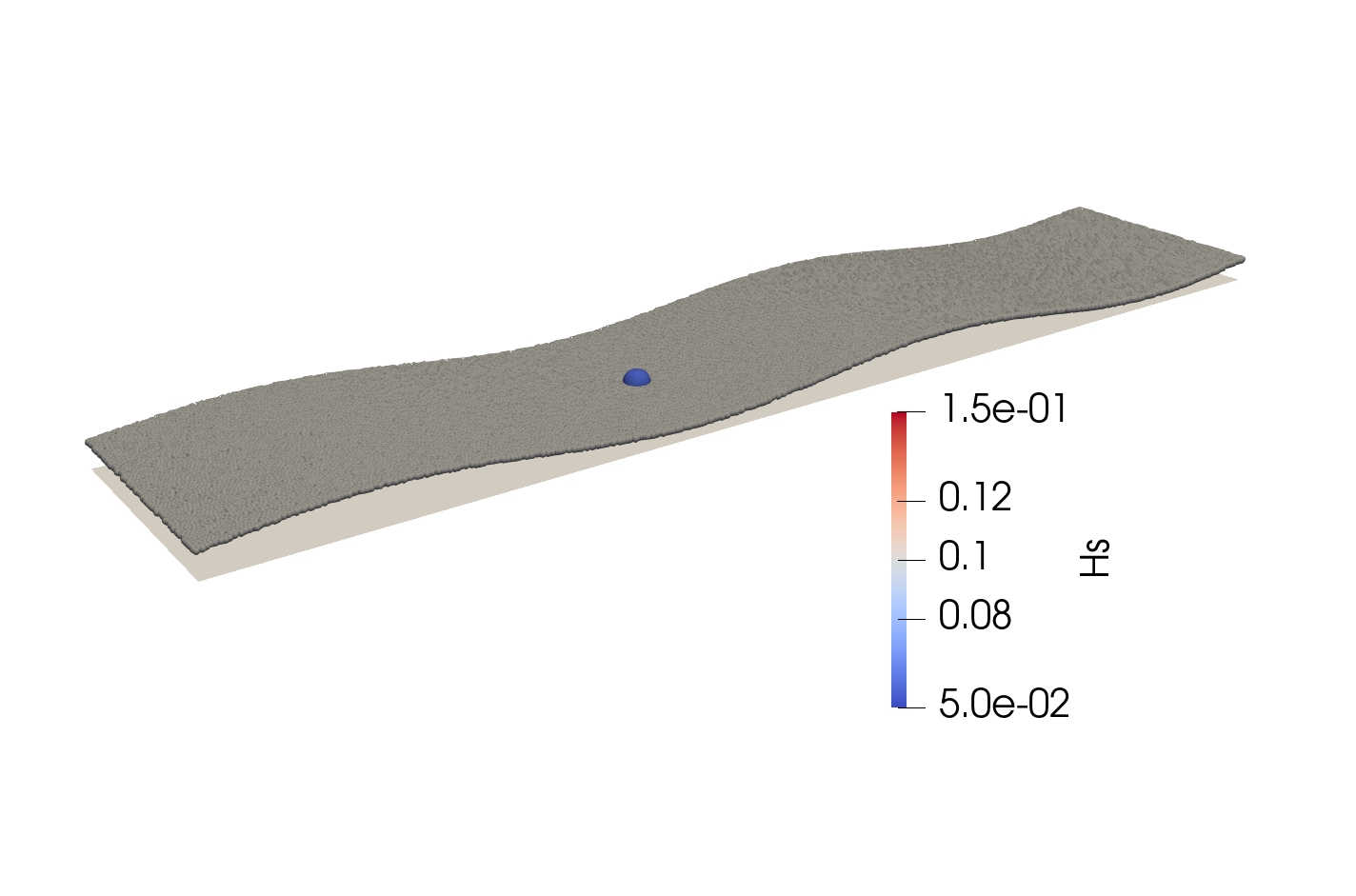}
    }\\
    \subfloat[t = 0.24s]{
    \includegraphics[width=0.5\textwidth]{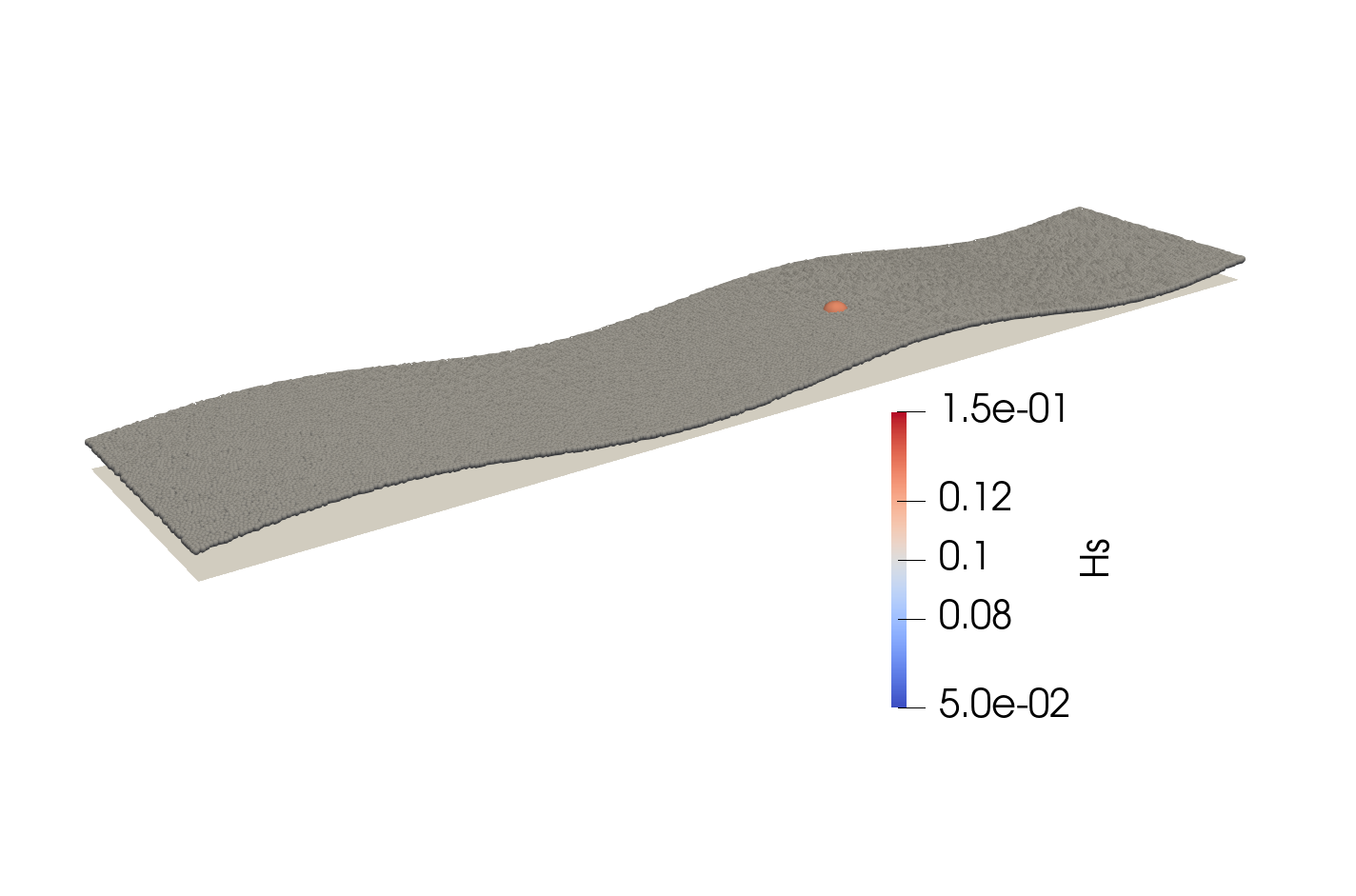}}
    \subfloat[t = 0.27s]{
    \includegraphics[width=0.5\textwidth]{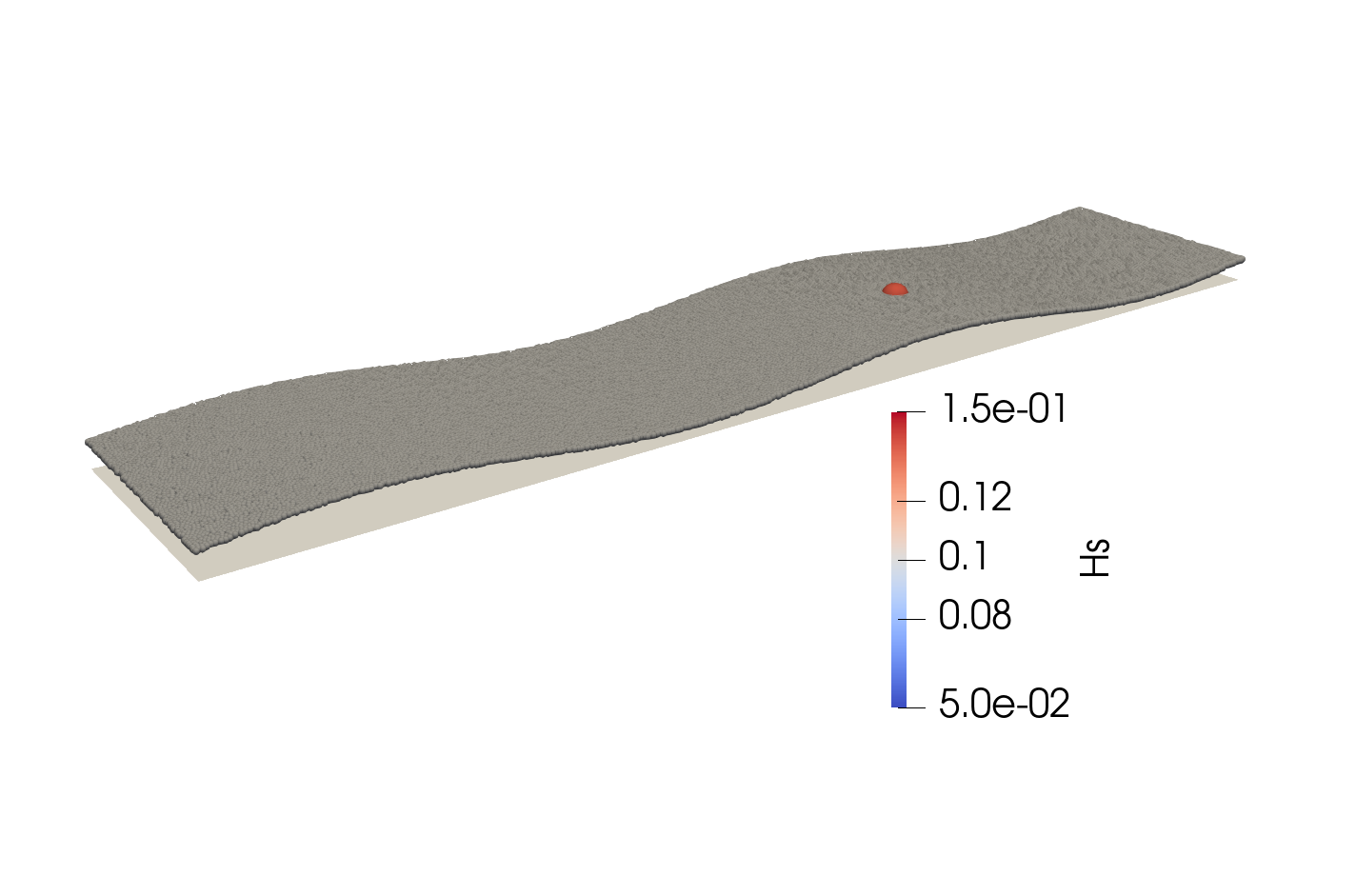}
    }
    \caption{Test Case 1: A droplet is tracked as it slides in the liquid thin-film over the wavy solidified region. The colour of the droplet represents its stored solid height. Through frames (a)-(f), the droplet is seen following the topology of the solidified region. As it moves, the solid height stored is higher in the crests (red colour) and lower in the troughs (blue colour). This is achieved by treating solid height as an advected property as per Eq.~\ref{eq:solid_advection}. Only one droplet is shown for easier visualization. }
    \label{fig:TC1_frames}
\end{figure}

\begin{figure}
    \centering
    \subfloat[]{
    \includegraphics[scale=0.425]{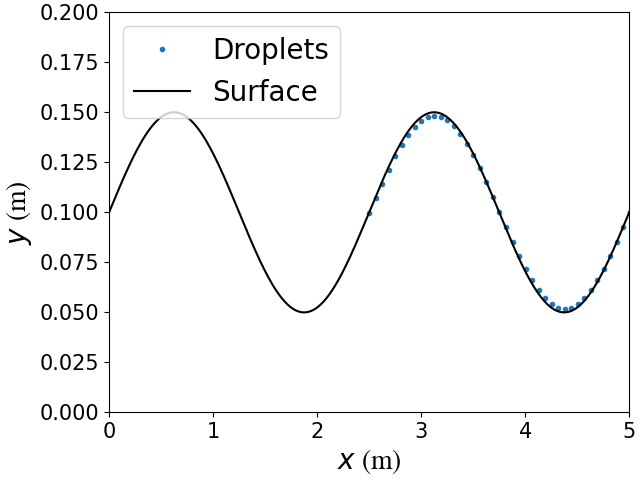}}
    \subfloat[]{\includegraphics[width=0.5\textwidth]{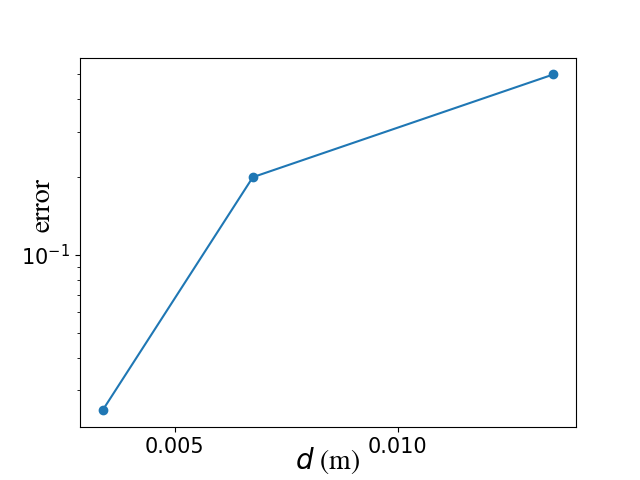}}
    \caption{\blue{Test Case 1: Error quantification: (a) Droplets positions at $t=0.25$s for $d_1$=3.375e-3m (b) Error convergence w.r.t droplet diameter;  an error of 2.5\% is obtained for the case with finest diameter ($d_1$).}}
    \label{fig:T1_error}
\end{figure}

In this test case, we consider a stationary, wavy solidified region over which the droplets of the thin-film slide at a constant velocity of $10$m/s magnitude. The purpose of this test case is to verify the movement of droplets over the solid and the advection of solid-height as the movement occurs. A flat surface, as shown in Fig.~\ref{fig:TC1_domain}, is considered and a  solidified region is assumed, whose height is given by
\begin{equation}
    H_s = 0.1 + 0.05 \text{sin}\left( \frac{4 \pi x }{L}\right)
    \label{eq:TC1}
\end{equation}
\blue{This equation represents the analytical solid height i.e. the wavy solid region. $L$ is taken to be 5 units.}
Fig \ref{fig:TC1_frames} shows a droplet of the thin-film being tracked as the film slides. The other droplets are hidden to enhance visualization.  The coloured contour of the droplet represents the solid height $H_s$ stored at its location. In Fig.~\ref{fig:TC1_frames}(a) and (b), the droplet is located close to a crest. The solid height $H_s$ is correspondingly high, as seen by the red colour of the droplet. As the droplet continues moving, it reaches a trough and takes on lower values of $H_s$, shown by the blue colour in Fig.~\ref{fig:TC1_frames}(c) and (d). The trend repeats in Fig.~\ref{fig:TC1_frames}(e) and (f), as the droplet goes on to climb the next crest. Therefore, it is seen from the frames in the figure that as the thin-film slides, the liquid droplet moves in accordance to the topology of the solidified region and also, updates the value of the solid height as per Eq.~\ref{eq:solid_advection}. \blue{ Since Eq.~\ref{eq:solid_advection} is an advection equation, it is likely to introduce numerical errors. The error is quantified and a convergence with respect to droplet diameter is reported. We consider three cases with different diameters such that $d_1$=3.375e-3, $d_2$ = 2$d_1$ and $d_3$ = 2$d_2$. In these three cases, the height of the liquid film remains the same, which implies that smaller diameters would correspond to a higher number of droplets. It is, therefore, evident that cases with smaller diameter of droplets will result in smaller numerical errors in comparison to cases with larger droplet diameters.  Fig.~\ref{fig:T1_error}(a) shows the position of droplets moving over the surface at $t=0.25$s, for the case with finest diameter of droplets ($d_1$). \green{It is noted that some droplets are located below the analytical surface, which is attributed to the dissipation errors due to the numerical solution of the advection of solid-height. } Fig.~\ref{fig:T1_error}(b) shows the convergence of error with droplet diameter. An error of 2.5\%, as defined in Eq.~\ref{eq:T1_error}, was obtained for the case with finest diameter. 

\begin{equation}
    \text{error} = \frac{1}{A_s}\int|H^{adv}_s-H_s|dA
    \label{eq:T1_error}
\end{equation}
Here, $H^{adv}_s$ denotes the solid height obtained from the advection equation, $H_s$ denotes the actual solid height in Eq.~\ref{eq:TC1} and $A_s$ denotes the area of the surface.}

\subsection{Test Case 2: Sliding thin-film of constant height over a solid layer of varying height}
\begin{figure}
    \centering
    \subfloat[A sketch of the solid profile given by Eq.~\ref{eq:TC2}, over which  droplets slide as governed by the DDM.]{
    \includegraphics[trim=0cm 6cm 0cm 8cm, width=0.55\textwidth]{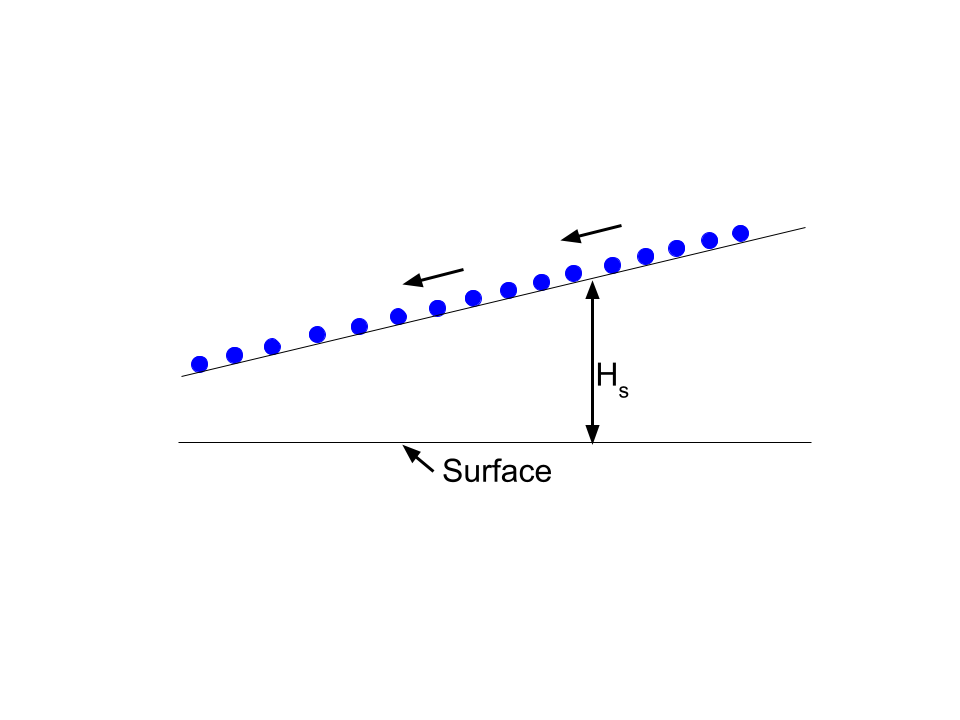}
    }
    \quad
    \subfloat[The numerical domain of computation: a flat plate initialized with an inclined solidified region.]{
    \includegraphics[trim=0cm 6cm 0cm 10cm, width=0.4\textwidth]{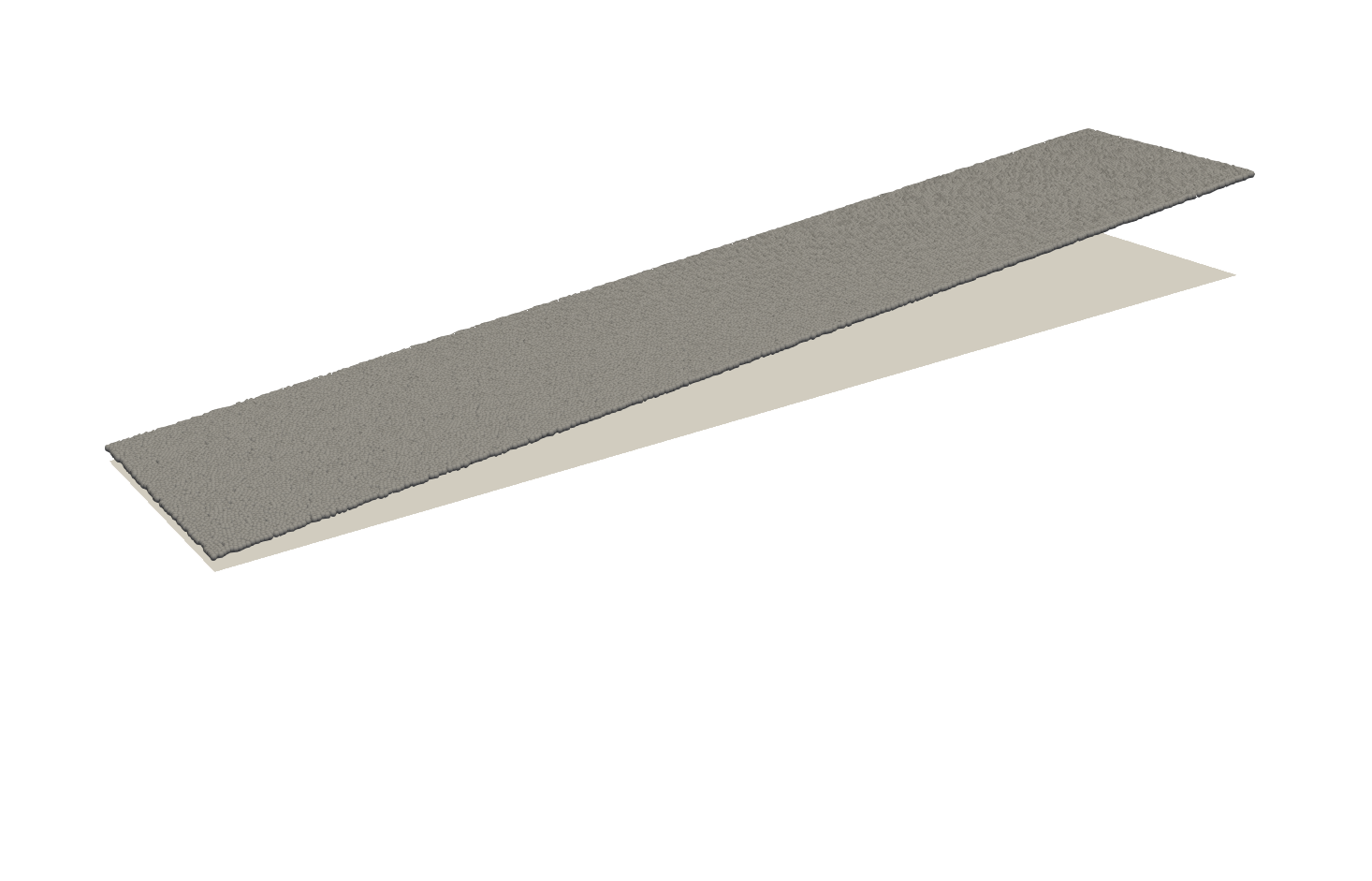}}
    \caption{Test Case 2: Domain of computation. }
    \label{fig:TC2_domain}
\end{figure}

In this test case, the DDM model that solves the liquid thin-film flow is verified in the presence of a solidified region of linearly varying height. As seen in Fig.~\ref{fig:TC2_domain}, a linearly varying solid height is considered. 
\begin{equation}
    H_s(x) = 0.05 + x \tan \left(\frac{\pi}{36}\right) 
\label{eq:TC2}
\end{equation}
This solid profile effectively creates a ramp of 5 degrees on which the liquid thin-film slides. The below analytical expression for velocity of a thin-film of constant height, can be used for comparing the velocity obtained numerically. 
\begin{equation}
    V_{\text{drop}} = g \sin \theta \frac{\rho H_l^2}{\eta}\left(1-\exp\left[-\frac{\eta t}{\rho H_l^2}\right]\right)
    \label{eq:TC2_analytical}
\end{equation}
where $V_{\text{drop}}$ is the magnitude of the velocity of the droplet in the direction along the solid ramp, \blue{$g = \| \vec{g} \|$ (=$10^3 m/s^2$) is the acceleration due to gravity, $\theta$ is the angle of inclination of the plane set to $5^o$, $\rho$ (=$1000$ kg/m$^3$) is the density of the liquid film, $\eta$ (=$0.001$ Ns/m$^2$) is the dynamic viscosity of the liquid film, and $H_l$ is the height of the liquid film. The droplets on the surface are distributed uniformly over the surface to achieve a film height $H_l \approx 3.6e-4$m.}  
\begin{figure}
    \centering
    \subfloat[]{
    \includegraphics[width=0.5\textwidth]{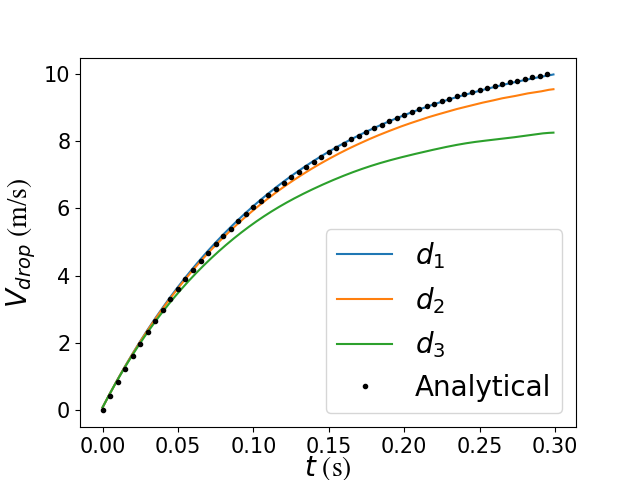}}
    \subfloat[]{
    \includegraphics[width=0.5\textwidth]{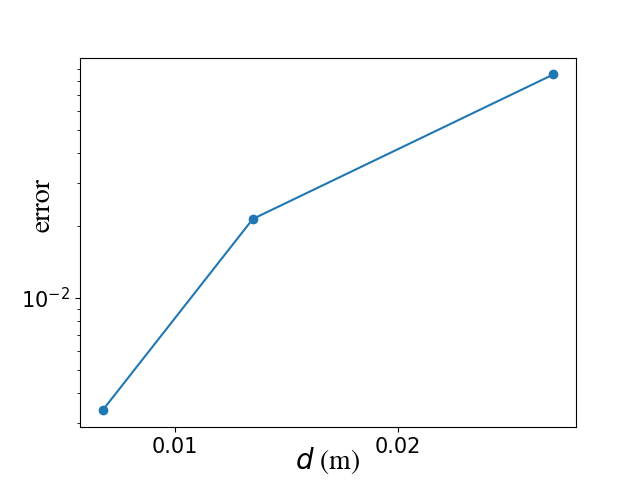}
    }
    \caption{\blue{Test Case 2: Comparison of velocity from the solution of the DDM model with Solidification (DDM-S) with the analytical solution; (a) Comparison of simulations with three different diameters,$d_1 = 6.75e-3$,$d_2 = 2d_1$, $d_3 = 2d_2$. (b) Error convergence plot. }}
    \label{fig:TC2_velocity_plot}
\end{figure}
\blue{As per Eq.~\ref{eq:TC2_analytical}, the liquid film starts from rest and accelerates under the influence of  gravitational forces until it reaches a terminal velocity due to the balancing viscous forces.} \\
\blue{ We present a convergence study by considering three cases of varying droplet diameters such that the height of the liquid film remains the same. The case with the smallest diameter corresponds to $d_1 = 6.75e-3$m and the other two cases correspond to $d_2 = 2d_1$ and $d_3 = 2d_2$. In order to ensure that these three cases have the same liquid film-height, the number of droplets are scaled accordingly i.e. the case with small diameter of droplets would have a higher number of droplets to maintain the same height of liquid film. } Fig.~\ref{fig:TC2_velocity_plot}(a) shows the evolution of velocity at a fixed point on the ramp. It compares the velocity obtained from the three simulations of DDM with solidification (DDM-S) with that of the analytical solution. \blue{The DDM-S solutions converge as the diameter of the droplets is refined. This is also quantified by the plot of error in Fig.~\ref{fig:TC2_velocity_plot}(b), in which the error is defined as 
\begin{equation}
    \text{error} = \frac{1}{T V_t}\int |V_{\text{DDMS}}-V_{\text{ana}}|dt
\end{equation}
where $V_{\text{DDMS}}$ is velocity obtained from the DDM-S model, $V_{\text{ana}}$ is the analytical solution, $T$ is the total simulation time and $V_t$ is the analytical terminal velocity of the film. For the case with the smallest diameter ($d_1$), the error is within 0.34\% of the analytical solution.}
\subsection{Test Case 3: Solidification of a stationary thin film }
\begin{figure}
    \centering
    \subfloat[A sketch of the domain: Stationary droplets are nearly uniformly distributed over a flat plate and the solidification occurs in the direction shown. ]{
    \includegraphics[trim=0cm 6cm 0cm 10cm, width=0.55\textwidth]{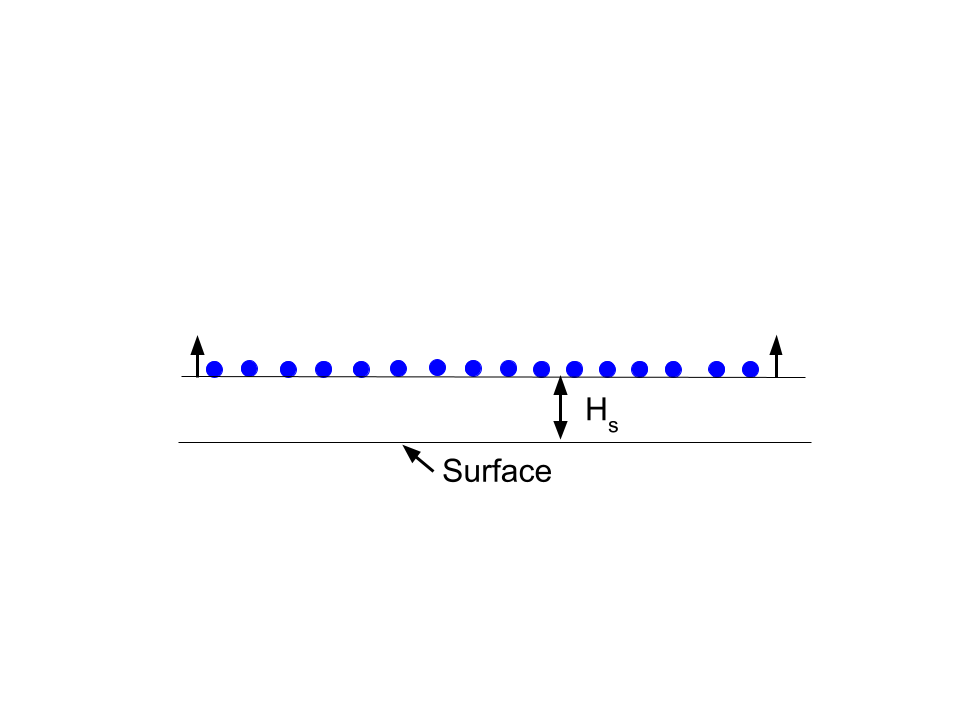}
    }
    \quad
    \subfloat[The numerical domain of computation: a flat plate with droplets initialized with $H_s=0$.]{
    \includegraphics[trim=0cm 6cm 0cm 10cm, width=0.4\textwidth]{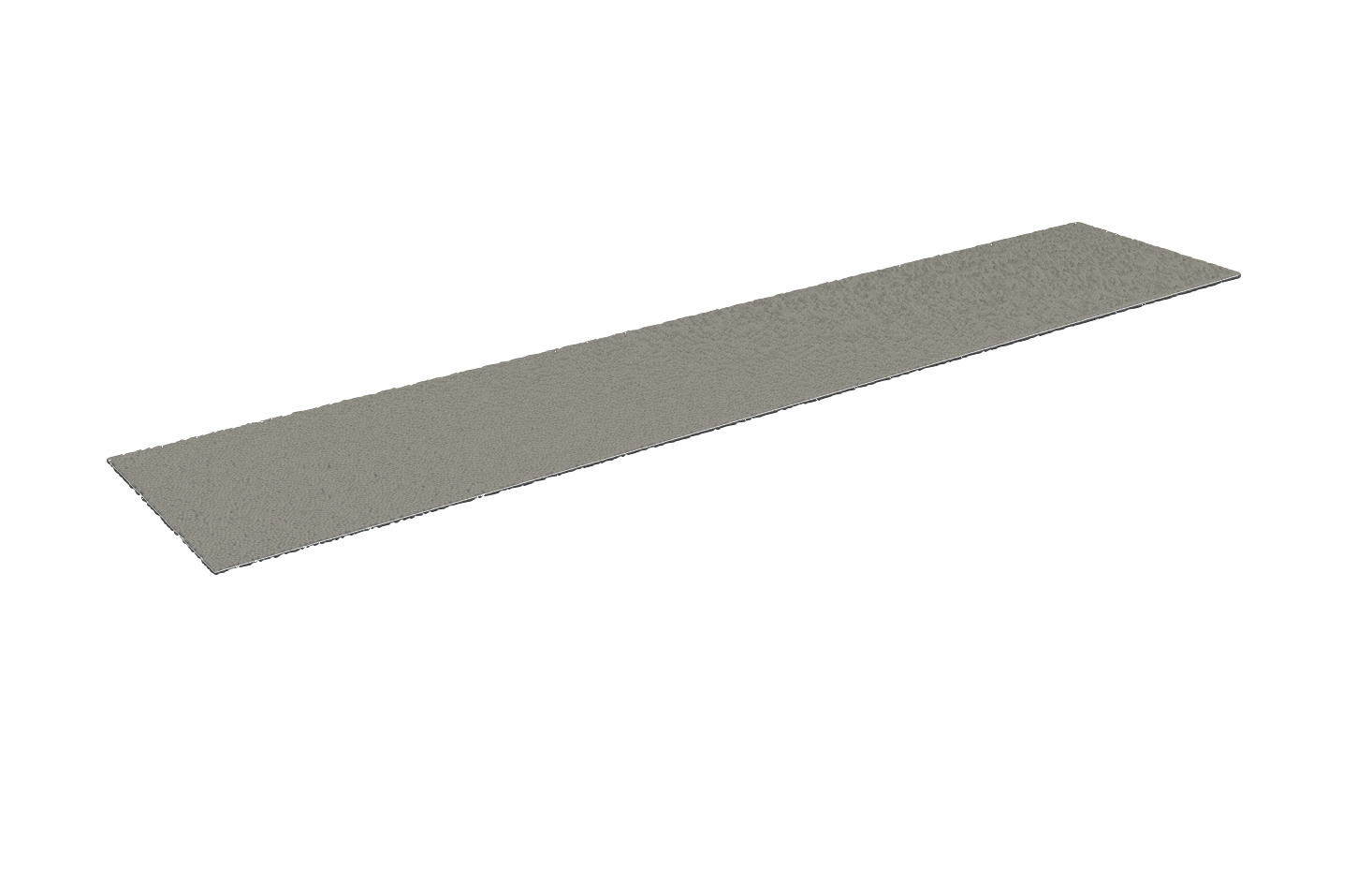}}
    \caption{Test Case 3: Domain of computation. }
    \label{fig:TC3_domain}
\end{figure}

\begin{figure}
    \centering
    \subfloat[Comparison of solid-liquid interface location]{
    \includegraphics[width=0.45\textwidth]{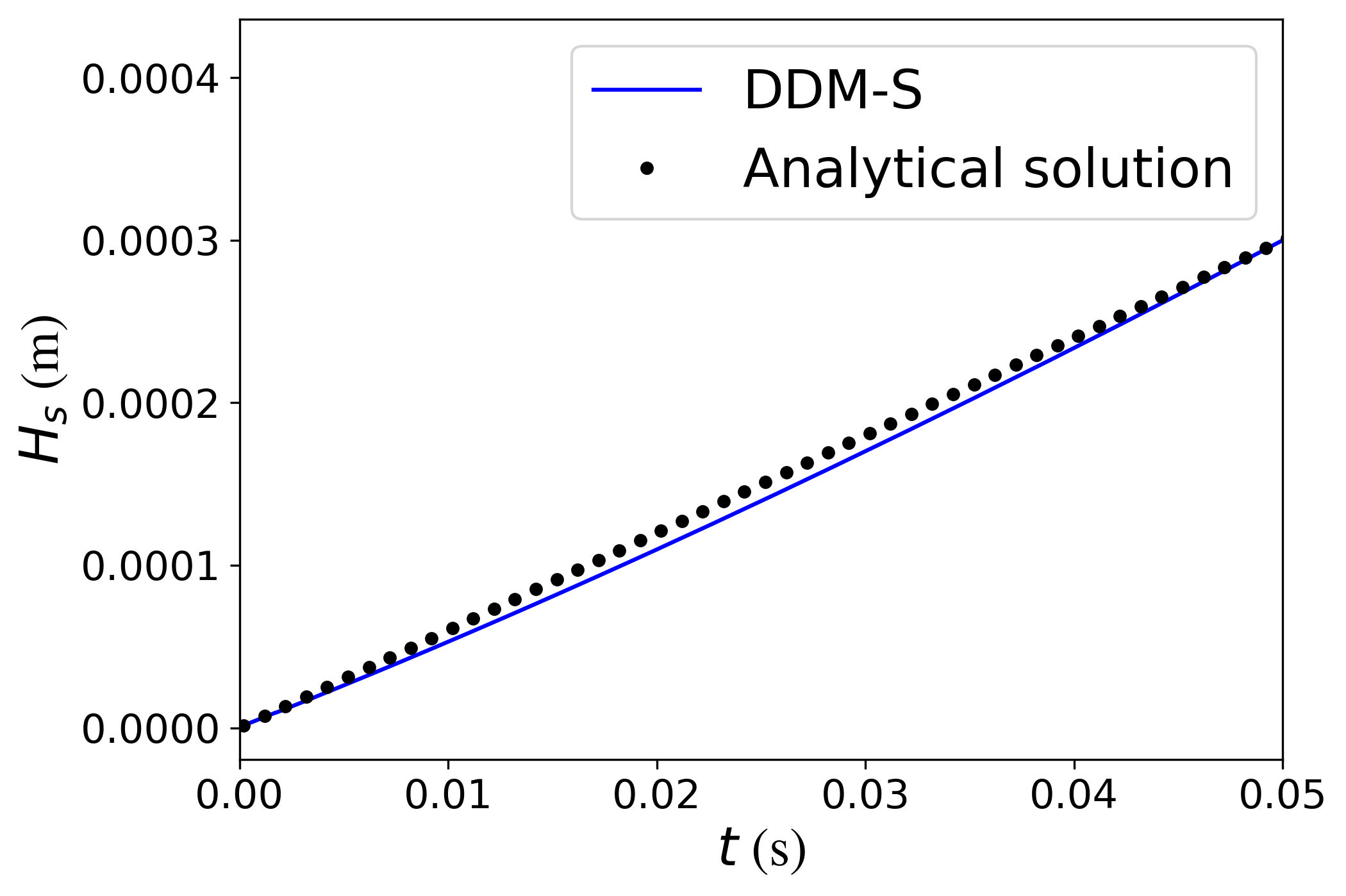}
    }
    \quad
    \subfloat[Comparison of temperature profiles]{
    \includegraphics[width=0.45\textwidth]{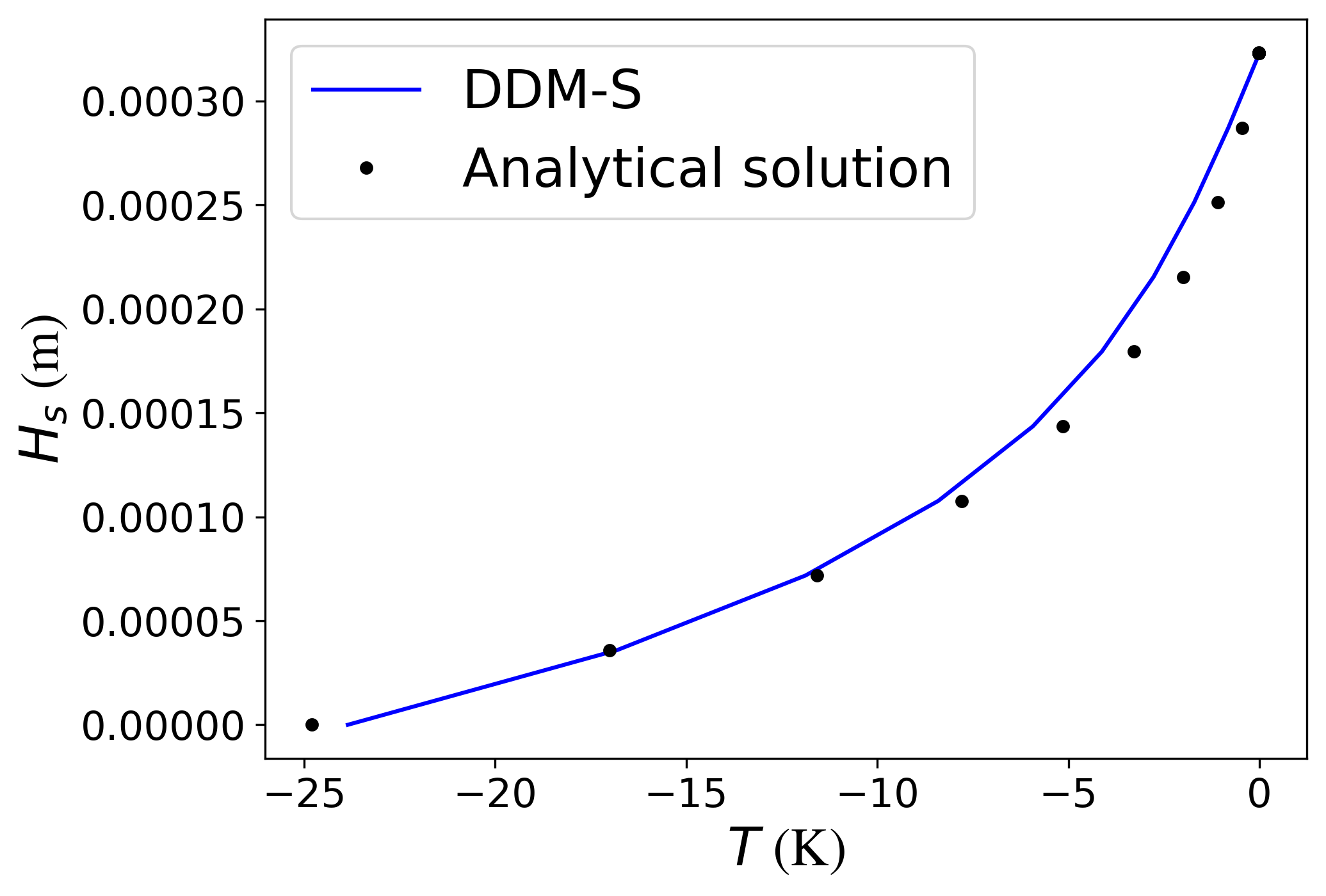}}
    \caption{Test Case 3: Comparison of numerical solution of DDM-S and analytical solution}
    \label{fig:TC3_results}
\end{figure}

\begin{figure}
    \centering
    \includegraphics[width=0.5\textwidth]{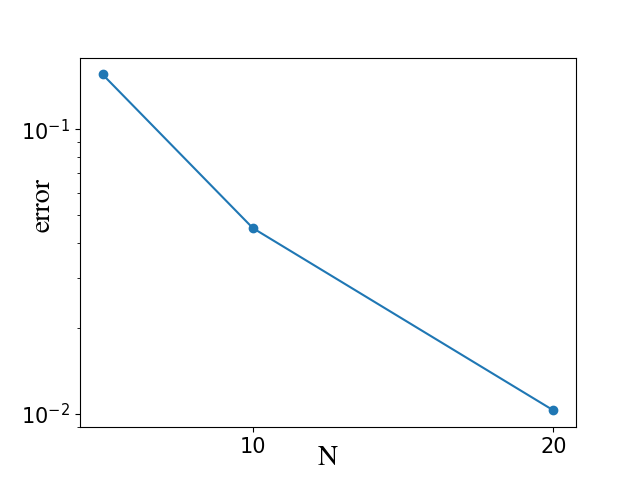}
    \caption{\blue{Test Case 3: Convergence plot. The error drops with the number of points discretizing the 1-D subdomain (N).}}
    \label{fig:T3_error}
\end{figure}

In this test case, a stationary film resting on a flat surface undergoes directional solidification starting from the surface, as shown in Fig.~\ref{fig:TC3_domain}. The temperature in the liquid film is uniformly at the melting temperature. The temperature in the solidified region varies from a forcing temperature at the surface to the melting temperature at the solid-liquid interface. An analytical solution for the interface location and the temperature profile is shown below \cite{stolze2016directional}. 
\begin{equation}
    H_s = v_I t 
\end{equation}
\begin{equation}
    T(n,t) = T_m + \frac{L}{C_v}\left[ 1 - \exp\left( \frac{v_I^2}{\alpha}t - \frac{v_I}{\alpha}n \right) \right]
    \label{eq:TC3_Tanalytical}
\end{equation}
\blue{Here, $L$ is the latent heat (=$10^3$J/kg), $C_v$ is the specific heat (=$10^3$J/kg), $v_I$ is the interface speed, $\alpha$ is the thermal diffusivity (=0.5m$^2$/s), $T_m$ (=0$^o$C) is the melting temperature and $n$ is the normal coordinate.} In the numerical simulation, a Dirichlet boundary condition is imposed at the surface that can be obtained by substituting $n=0$ in Eq.~\ref{eq:TC3_Tanalytical}. Imposition of this boundary condition, pushes the solid-liquid interface away from the surface causing a growth in the solid height ($H_s$).
Fig.~\ref{fig:TC3_results}(a) compares $H_s$ of the DDM-S model with the analytical solution and Fig.~\ref{fig:TC3_results}(b) compares the temperature profiles of the analytical and numerical solution at $t=0.07$s. \blue{These results correspond to 20 points discretizing the 1-D sub-domain. Fig.~\ref{fig:T3_error} shows the convergence of error with respect to the number of points ($N$) used to discretize the 1-D sub-domain. Beyond $N$=20, the solution does not change significantly.}

\subsection{Test Case 4: Solidification on a circular conductor with droplet impingement}
Having verified the different components of the proposed thin film solidification model in the previous test cases, we now validate the DDM-S model as a whole. 
In this test case, a circular, cylindrical conductor is considered on which water droplets impinge to form ice. At the beginning of the simulation, there is no water film or ice present on the surface. The icing conditions are tabulated in Table.~\ref{tab:TC4: Icing conditions}. 
\begin{table}[]
    \centering
    \caption{Test Case 4: Test conditions}
    \begin{tabular}{cc}
    \hline
       $V_\infty$  & 10$m/s$ \\
        $T_a$, $T_s$, $T_{avg}$ & $-5^o$ \\
        Liquid Water Content (LWC) & $1.8\text{g/m}^3$\\
        Mean Volume Diameter & 26 $\mu$m\\
        Icing time & 10 minutes \\
        Cylinder diameter & 34.9 mm \\
         \hline 
    \end{tabular}
    \label{tab:TC4: Icing conditions}
\end{table}
Fig.~\ref{fig:CylinderGeometry} shows the geometry of the circular cylinder. A fictitious plane upstream of the geometry is initialized with randomised positions of droplet sources. The rate of droplet generation depends on the volume flux of water through this plane. The volume flux of water droplets through the plane of area $A$, is
\begin{equation}
    Q_w = Q_{a} \frac{\text{LWC}}{\rho_w} = A V_\infty \frac{\text{LWC}}{\rho_w} 
    \label{eq:volume_flux}
\end{equation}
The subscripts $w$ and $a$ denote water droplets and air respectively, $\rho$ denotes density, $\text{LWC}$ stands for Liquid Water Content and $V_{\infty}$ denotes the freestream air velocity. The diameter of the droplets is related to the number of droplets introduced per second, $N_{d}$, at the fictitious plane, as
\begin{equation}
    N_d \frac{\pi}{6}d^3 = Q_w 
    \label{eq:numberofdroplets}
\end{equation} 
The above equations, Eqns.~\ref{eq:volume_flux} and \ref{eq:numberofdroplets}, are used in test cases which involve impinging droplets with a given LWC, as done in the next test case too. \blue{In the numerical model, the heat transfer coefficients at the interfaces of the thin-film with the ambient and the surface, $h_s$ and $h_a$, are chosen to be $100$ W/m$^2$K.} \\

{\blue{Fig.~\ref{fig:TC4_frames} shows the development of the ice profiles. From Fig.~\ref{fig:TC4_frames}(a)-(d), an increase in $H_s$ is noted by the darkening red coloured contours of $H_s$. The increase in $H_s$ can also be noticed as a widening gap between the droplets and the cylinder surface.} The numerical solution of the model DDM-S, is obtained by interpolating the solid (ice) heights onto the mid-section of the cylindrical conductor. \blue{Fig.~\ref{fig:CylinderGeometryResutls} compares this interpolated ice shape, with that reported in literature\cite{chao2021glaze}, where a glaze ice model was used to predict the icing. We define an error as in Eq.~\ref{eq:error}. We note an error of around 11 \%  with respect to the numerical data reported in literature.}
\blue{
\begin{equation}
    \text{error} = \frac{\int_S |H_{s,DDM-S}-H_{s,lit}|}{\int_S H_{s,lit}}
    \label{eq:error}
\end{equation}}
\begin{figure}
    \centering
    \includegraphics[trim={5cm 0 5cm 0},width=0.5\textwidth]{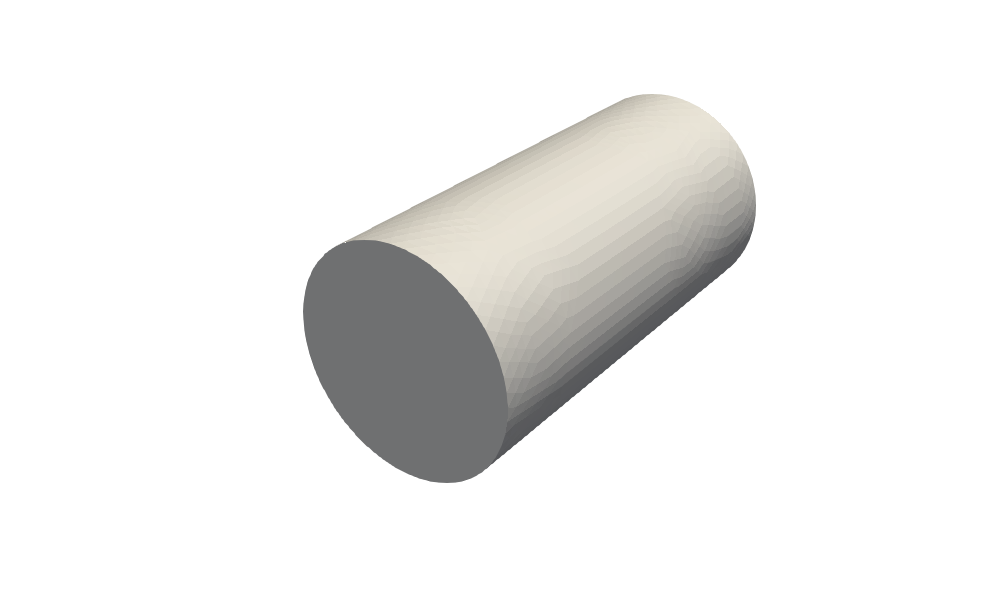}
    \caption{Test Case 4: Geometry of circular conductor.}
    \label{fig:CylinderGeometry}
\end{figure}
\begin{figure}
    \centering
    \subfloat[t/T = 0.25]{
    \includegraphics[trim={2cm 0 2cm 0},width=0.5\textwidth]{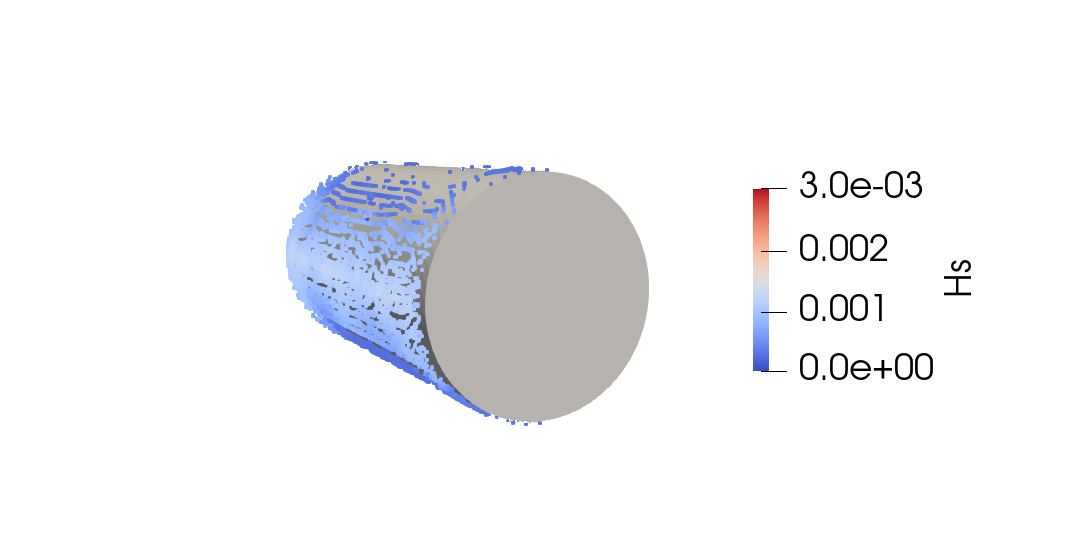}}
    \subfloat[t/T = 0.5]{
    \includegraphics[trim={2cm 0 2cm 0},width=0.5\textwidth]{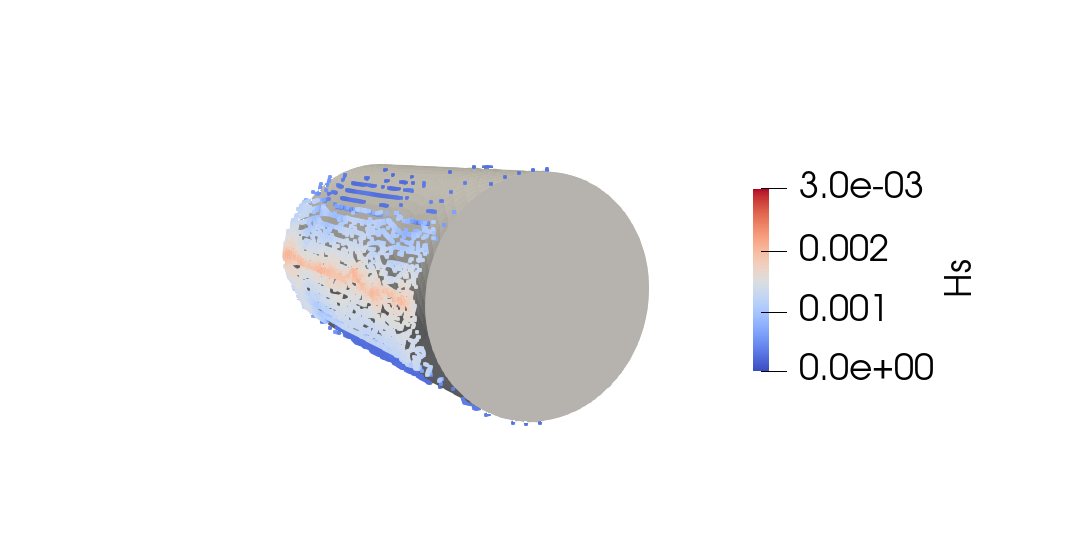}
    }\\
    \subfloat[t/T = 0.75]{
    \includegraphics[trim={2cm 0 2cm 0},width=0.5\textwidth]{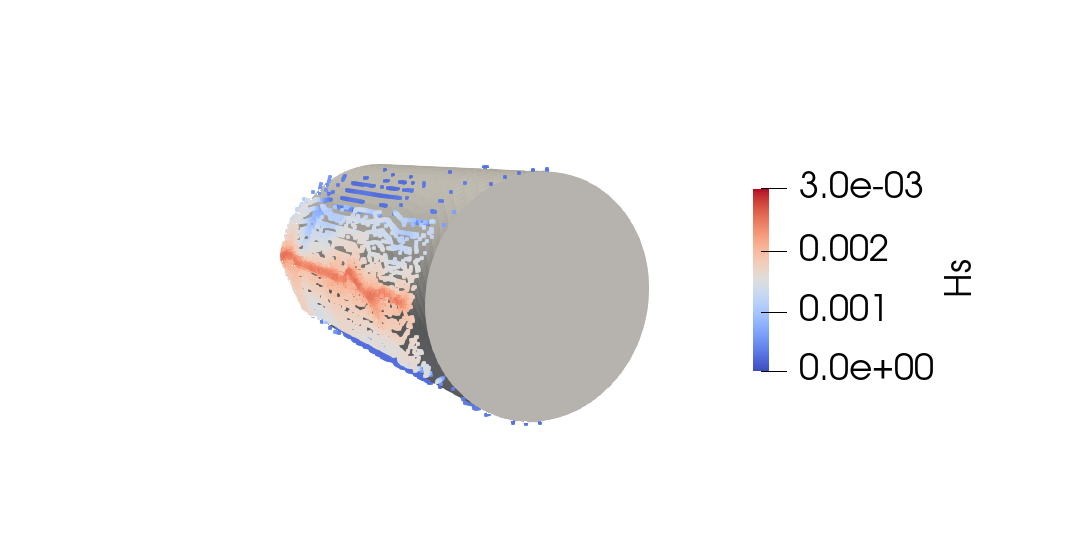}}
    \subfloat[t/T = 1]{
    \includegraphics[trim={2cm 0 2cm 0},width=0.5\textwidth]{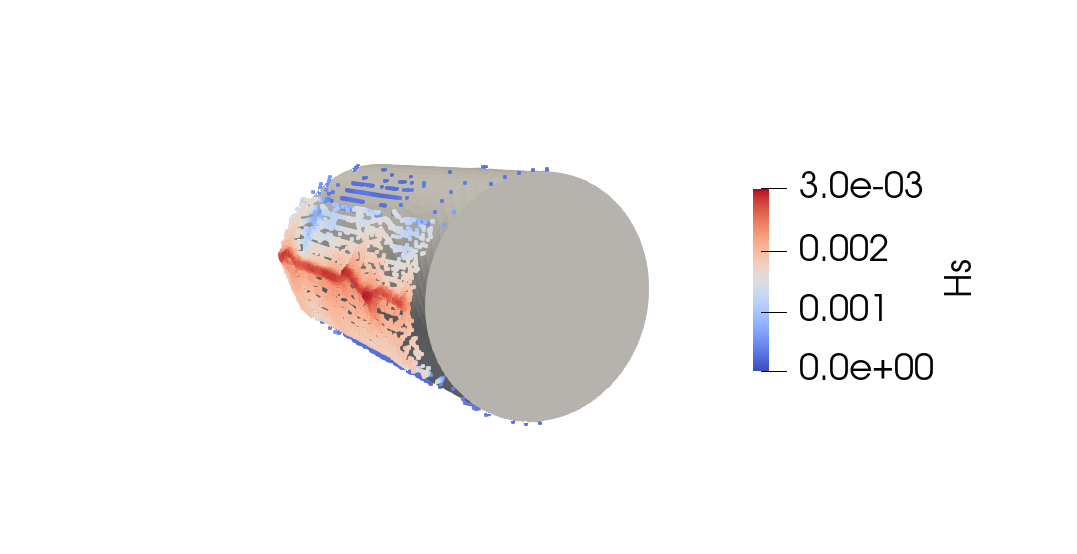}
    }\\
    \caption{\blue{Test Case 4: Formation of ice on a circular conductor. The coloured contours of ice-height show an increasing trend with simulation time.}}
    \label{fig:TC4_frames}
\end{figure} 
\begin{figure}
    \centering
    \includegraphics[,width=0.5\textwidth]{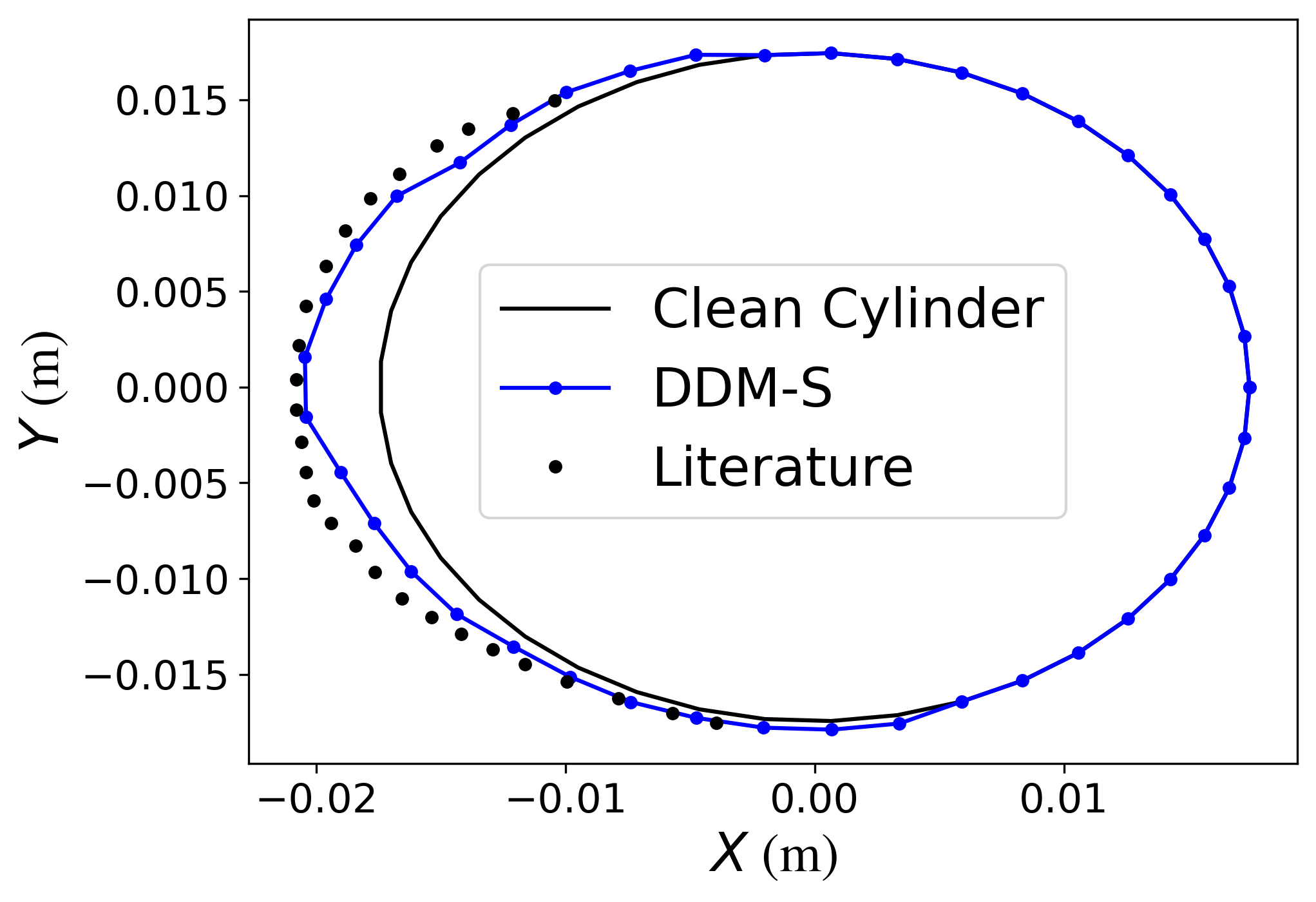}
    \caption{Test Case 4: Solidified region of ice on a circular cylinder; results compared with literature \cite{chao2021glaze}. }
    \label{fig:CylinderGeometryResutls}
\end{figure}

\subsection{Test Case 5: Solidification on an aircraft wing with droplet impingement}
In this test case, a  GLC-305 wing is considered with droplet impingement. The reference test case is taken from the experimental data of the Icing Research Tunnel (IRT) of NASA Glenn Research Centre \cite{papadakis2003aerodynamic}. The icing condition chosen for simulation corresponds to Ice-3 of the experimental data. Fig.~\ref{fig:GLC305} shows the wing geometry. The details of the wing are tabulated in Table.~\ref{tab:wingdetails}. 
\begin{figure}
    \centering
    \includegraphics[trim={10cm 0 10cm 0},clip,width=0.5\textwidth]{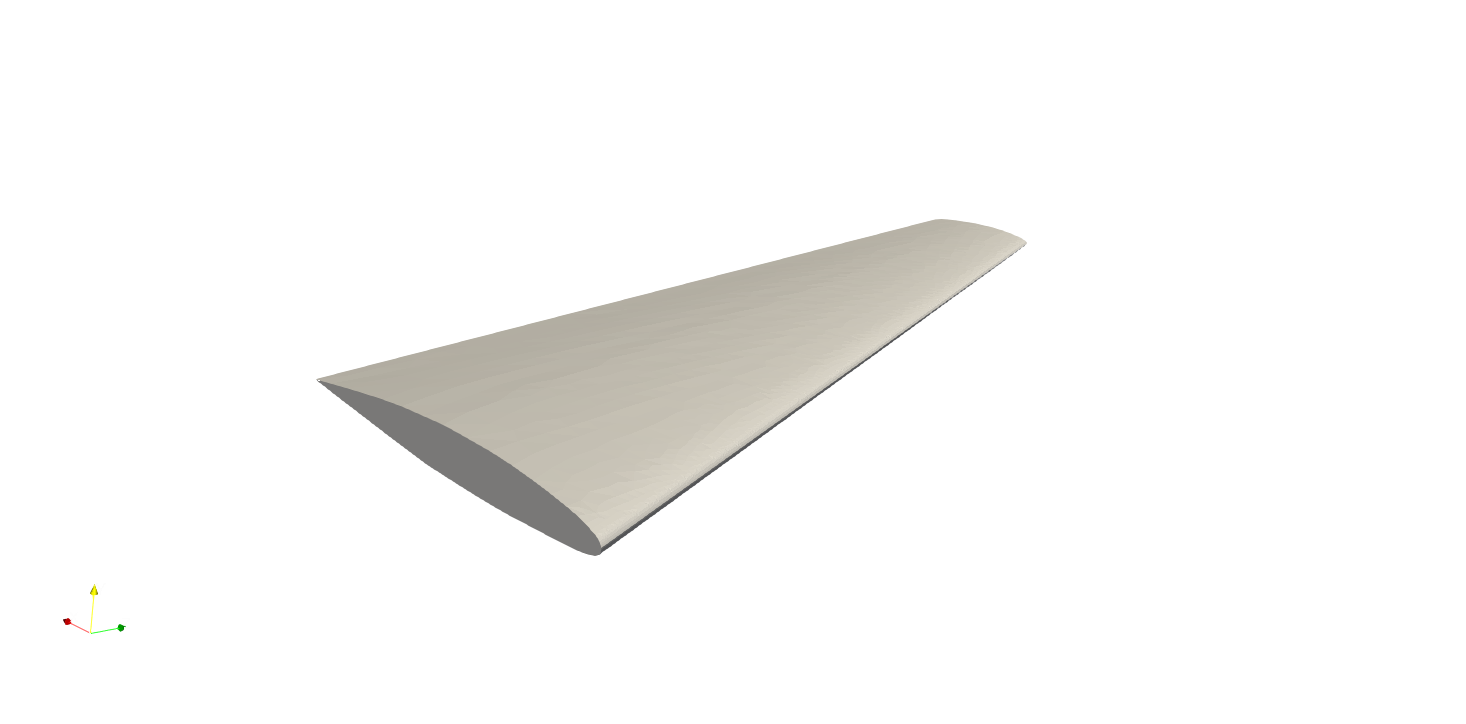}
    \caption{Test Case 4: GLC-305 wing geometry.}
    \label{fig:GLC305}
\end{figure}
\begin{table}
\centering
\caption{Test Case 3: Geometry and icing conditions}
\subfloat[GLC-305 wing details]{
    \begin{tabular}{cc}
       \hline 
       Airfoil Section  &  GLC-305\\
       Angle of attack  &  $6^o$\\
       Wing twist &  $4^o$\\
       Root chord & $25.2$ in\\
       Tip chord & $10.08$ in\\ 
       \hline 
    \end{tabular}
    \label{tab:wingdetails}
}
\hspace{2cm } 
\subfloat[Icing conditions]{
    \begin{tabular}{cc}
       \hline 
       $V_\infty$ &  89.7 m/s\\
       $T_\text{a}$,  $T_\text{s}$, $T_{\text{avg}}$ &  $-11.27^o$\\
       Liquid Water Content (LWC) &  $0.51\text{g/m}^3$\\
       Mean Volume Diameter & 14.5 $\mu$m\\
       Icing time & $5$ minutes\\ 
       \hline 
    \end{tabular}
    \label{tab:IC}
}
\end{table}
The conditions of icing are tabulated in Table.~\ref{tab:IC}. The numerical values of the parameters of the solidification model are derived from these icing conditions, as done in the previous test case, and with the help of inputs from the NASA Glenn Research Centre. \blue{In the numerical model, as in the previous test case, the heat transfer coefficients at the interfaces of the thin-film with the ambient and the surface, $h_s$ and $h_a$, are chosen to be $100$ W/m$^2$K.} \\
Fig.~\ref{fig:TC5_frames} shows the formation and evolution of the liquid thin-film (water) over the lower surface of the wing and its solidification to form ice. The coloured contours denote the ice height. It is noted that as the simulation time increases, the water thin-film develops over the wing surface. The ice-height at the leading-edge is higher than the other parts of the wing and it increases with time.   
\begin{figure}
    \centering
    \subfloat[t = 2s]{
    \includegraphics[width=0.5\textwidth]{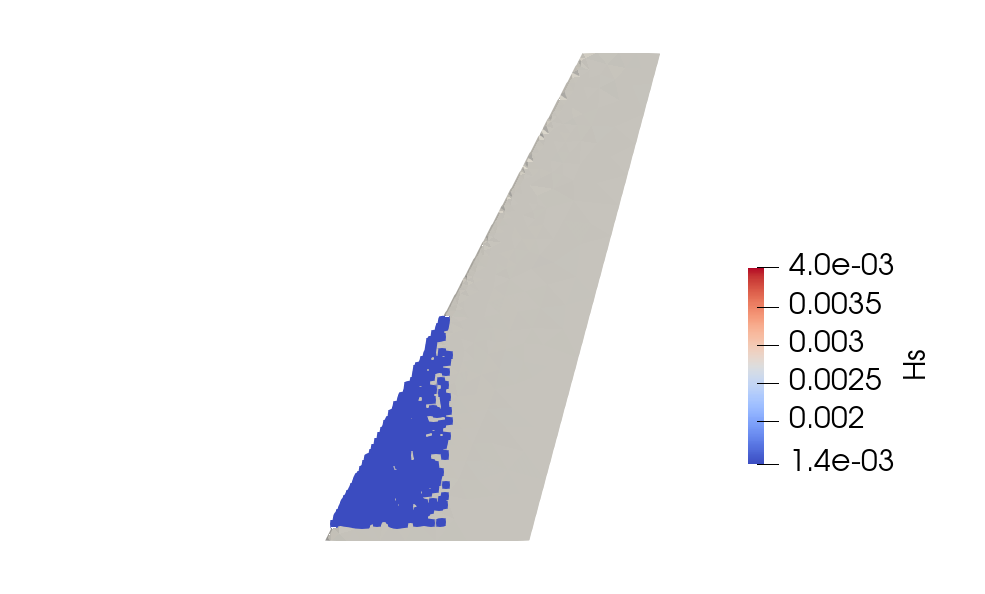}}
    \subfloat[t = 15s]{
    \includegraphics[width=0.5\textwidth]{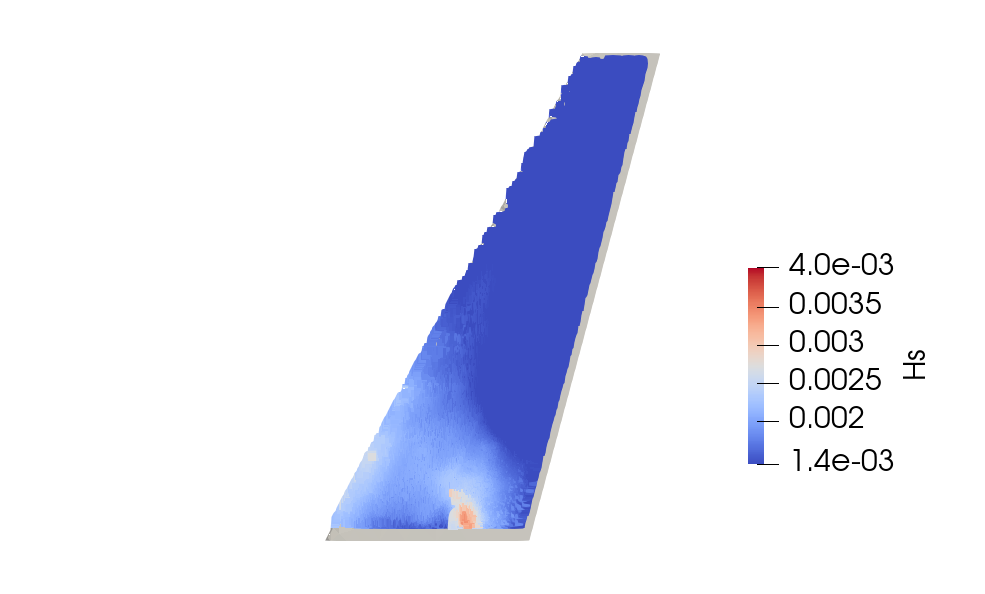}
    }\\
    \subfloat[t = 25s]{
    \includegraphics[width=0.5\textwidth]{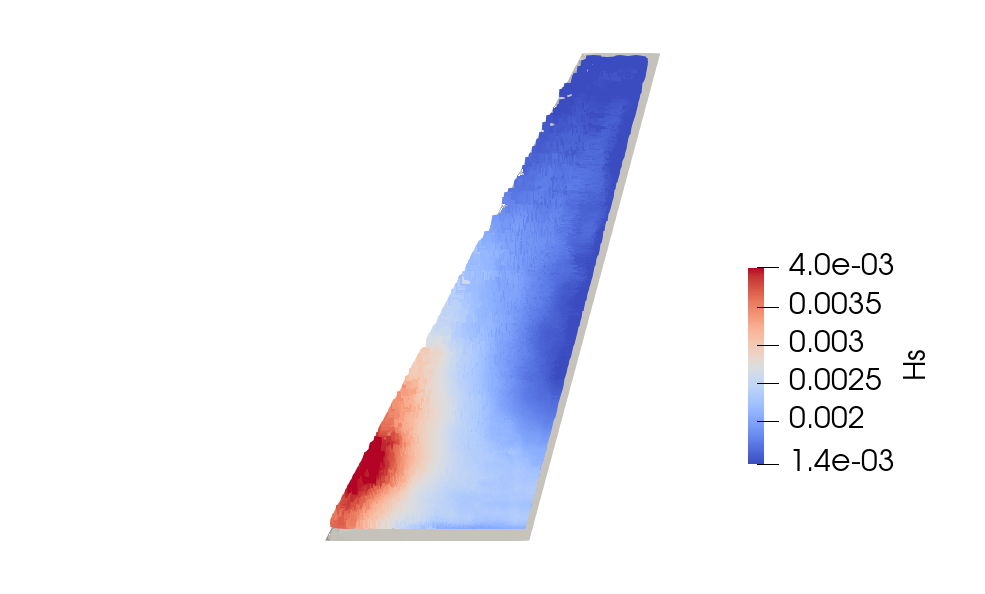}}
    \subfloat[t = 30s]{
    \includegraphics[width=0.5\textwidth]{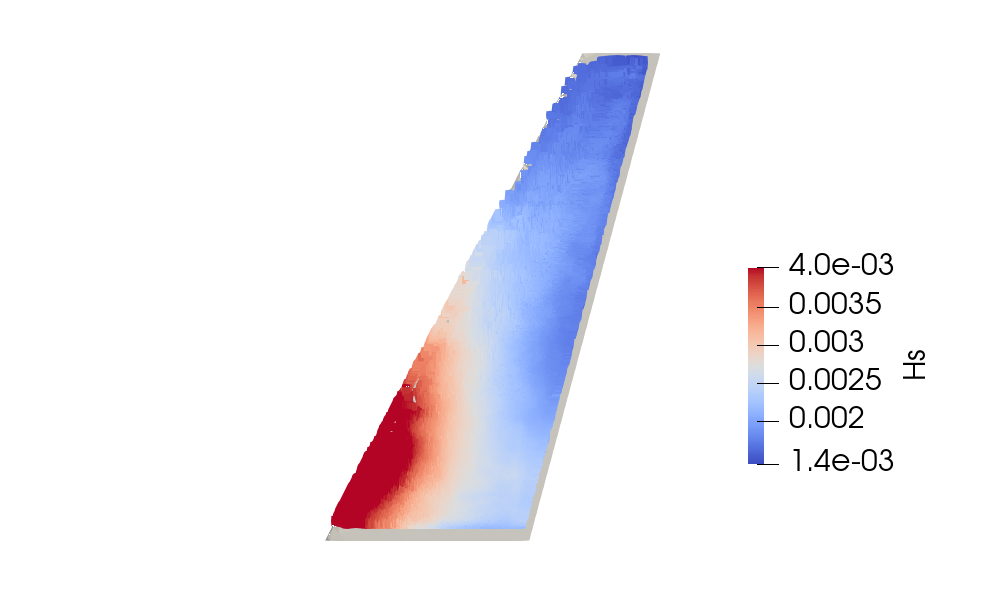}
    }\\
    \caption{Test Case 5: Formation of water film on the lower surface of the wing. The coloured contours of ice-height show an increasing trend with simulation time and larger ice accretion at the leading edge. }
    \label{fig:TC5_frames}
\end{figure}
Fig.~\ref{fig:WingIcingResults} shows the ice shape based on interpolated ice heights at the root section of the wing. The predictions of the proposed DDM-S model matches well with the experiment near the leading edge of the airfoil. However, further downstream of the leading edge, the model over-estimates ice-accretion in comparison to the experiment. \green{A possible source of this mismatch is that the aerodynamics plays a prominent role in droplet incidence on the wing, which has not been considered in the present model.} \green{The evolution of the thin-film, therefore, does not account for the shear and inertial forces exerted by the bulk-flow on the film, which is a limitation of the model in its present form. It is also noted that the liquid water content (LWC) used in the experiment may have been under-reported, as confirmed by NASA Glenn Research Centre. However, this effect is difficult to quantify.} 

\green{The role of aerodynamics will be actively captured in future extensions of the model which will involve the coupling of the present model with a bulk-flow model. For the purposes of the present work, its effect has been indirectly considered by using the collection efficiency $\zeta$ which signifies the amount of droplet incidence occurring at a given section on the surface \cite{cao2012numerical}. We consider a distribution of collection efficiency $\zeta$ over the airfoil using the freestream velocity vector ($\vec{V}_\infty$) and the outward surface normal ($\hat{n}$).
\begin{equation}
    \zeta = -\frac{\vec{V}_\infty \cdot \hat{n}}{|\vec{V}_\infty|}
\end{equation}
The distribution is shown in Fig \ref{fig:WingIcingResults2}(a).
The abscissa denotes the ratio of arc length from the leading edge (s) to the chord (c), with positive values signifying the upper surface and negative values signifying the lower surface. The collection efficiency is high close to the leading edge (s/c=0) and decays with distance from the leading edge. This decay is quicker over the upper surface than the lower surface due to the positive angle of attack. The mass of droplets impinging at a location is given by}
\green{
\begin{equation}
    \Dot{m}_{imp}  = \text{LWC } V_\infty \zeta A 
\end{equation}
}
\green{Here, $A$ denotes the area of impingement. At given locations that discretise the wing, we reproduce droplet impingement by placing droplet sources that generate the volume of water proportional to the collection efficiency. Thus, the change in volume of the droplets is estimated using the below relation
\begin{equation}
                \rho_w \frac{d \text{Vol}}{dt} = \Dot{m}_{imp} 
\end{equation}
where, $\rho_w$ and Vol denote the density of water and volume of a droplet respectively. Having indirectly modelled the effect of aerodynamics through the collection efficiency $\zeta$, the DDM-S solver proceeds to estimate the ice height. As seen in Fig.~\ref{fig:WingIcingResults2}(b), the ice heights show a relatively closer match with the experimental data in comparison to the previous case (Fig.~\ref{fig:WingIcingResults}) where $\zeta$ was not considered in the droplet impingement model. This result, therefore, emphasizes
the importance of coupling the aerodynamics of a bulk-flow with the DDM-S model.
}

\begin{figure}
    \centering
    \includegraphics[width=0.5\textwidth]{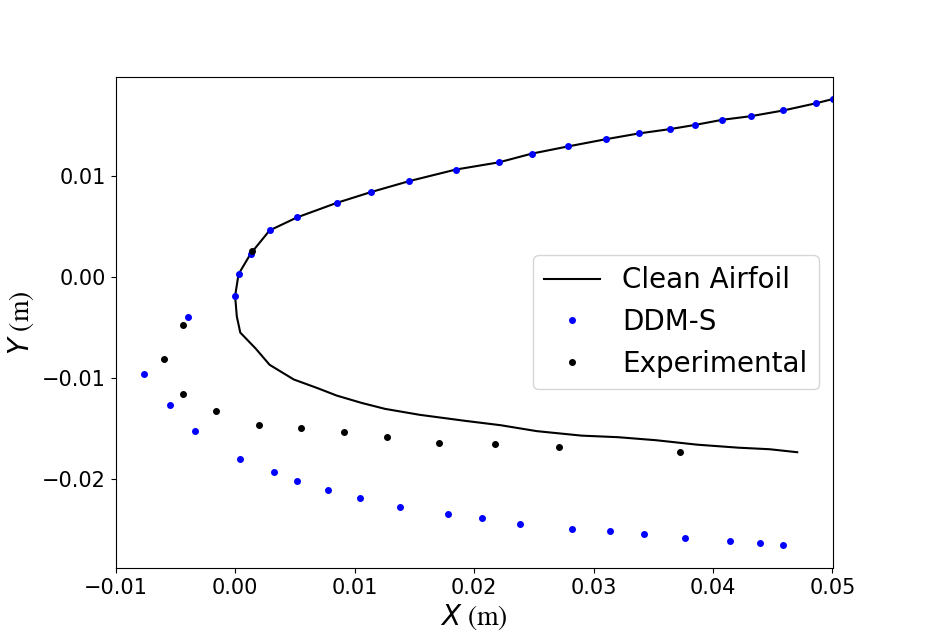}
    \caption{Test Case 5: Comparison of ice profiles at the root section obtained from the model with experimental data \cite{papadakis2003aerodynamic}.}
    \label{fig:WingIcingResults}
\end{figure}

\begin{figure}
    \centering
    \subfloat[]{
    \includegraphics[width=0.5\textwidth]{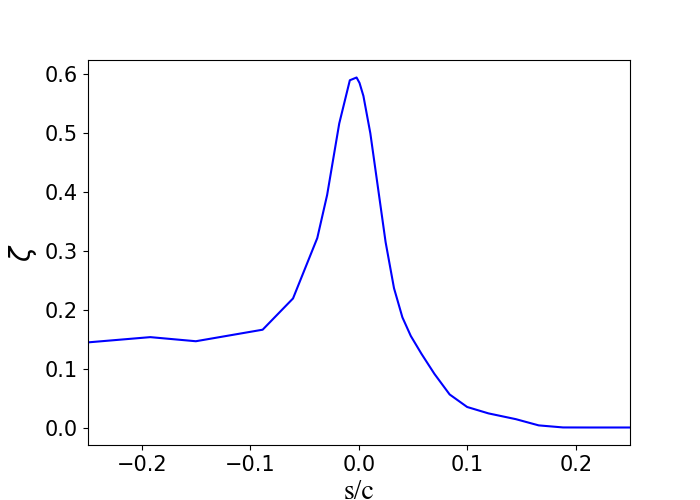}
    }
    \subfloat[]{
    \includegraphics[width=0.5\textwidth]{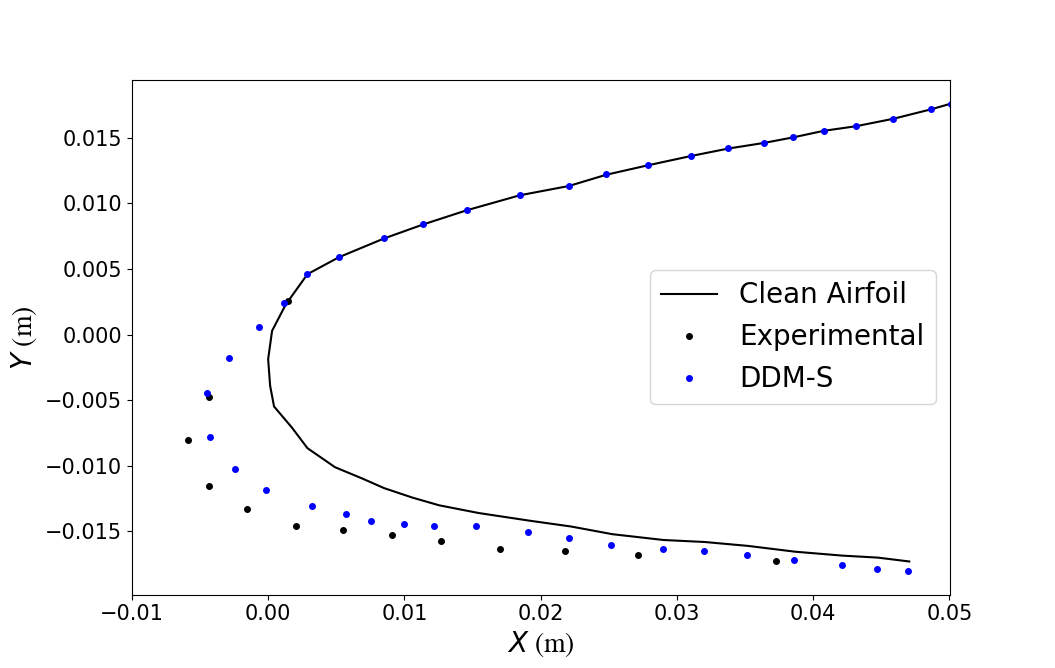}}
    \caption{\green{
Test Case 5: (a) Distribution of collection efficiency $\eta$ over the wing cross-section (b) Comparison of ice profiles at the root section obtained from the model with experimental data \cite{papadakis2003aerodynamic}.}}
    \label{fig:WingIcingResults2}
    \end{figure}
\section{Conclusions}
The Discrete Droplet Method with Soldification (DDM-S) is based on a previously proposed liquid thin-film model (DDM) that incorporates solidification phase-change. The numerical model stores solid height as a property at each droplet location. The evolution of the solid film thickness depends on the movement of the solid-liquid interface, as dictated by the Stefan problem. The movement of the droplet occurs over the solidified region and the velocity is governed by the momentum equation as in DDM. 

DDM-S accounts for unsteadiness in temperature distributions in the liquid and solid regions of the thin-film. It also allows the liquid temperature to evolve as per the heat transfer occurring with the surface without the assumption of perfect thermal contact. It also conveniently incorporates volumetric change in its framework. One of the unique aspects of the model is the treatment of the solid-height as a property at a droplet location. This treatment has some advantages and disadvantages. The advantage is that in this framework, the solid region can also have its own velocity and this velocity can be accounted for in the advection velocity of the solid-height advection equation, Eq.~\ref{eq:solid_advection}. This is done without a new set of numerical points, but simply using existing set of droplets as the numerical points. However, the disdvantage with this approach is that solid-height cannot be modelled without droplets and its evolution as per an advection equation makes it susceptible to numerical errors such as dissipation and dispersion, that can be countered only by using higher-order discretizations. 

Some of the future directions of research would include coupling the DDM-S with a bulk-flow Navier-Stokes model to make predictions in the presence of aerodynamic loads. This would be very useful in aircraft wing and wind turbine applications. Another interesting direction is to allow for the movement of the solid phase that could be governed by a different set of governing equations. This could help in problems like simulation of ice-sheet movement.  
\section*{Acknowledgements}
Anand S Bharadwaj would like to acknowledge Fraunhofer-Gesellschaft and SERB-NPDF for providing financial support during the course of the research. 
The authors would like to acknowledge Dennis Eck, Jr. of NASA Glenn Research Centre for providing valuable information about the test case of icing on GLC-305 swept wing. 
Pratik Suchde would like to acknowledge funding from the Institute of Advanced Studies, University of Luxembourg, under the AUDACITY programme. 

\appendix 
\section{Temperature profile at droplet location}
Let us consider the temperature profile in the liquid thin-film of the form \label{AppA}
\begin{equation}
    T(n) = a_l n^2 + b_l n + c_l
    \label{eq:quad}
\end{equation}
The constants $a_l$, $b_l$ and $c_l$ are determined by the conditions
\begin{equation}
    T(0) = T_m
    \label{c1}
\end{equation}
\begin{equation}
    \frac{\partial T}{\partial n} \Big|_{n=H_l} = -\frac{h_a}{\kappa}(T(H_l)-T_a) = \alpha
    \label{c3}
\end{equation}
\begin{equation}
    \frac{1}{H_l}\int_0^{H_l} T(n) dn = T_{avg} 
        \label{c4}
\end{equation}
Here, $n$ is the normal coordinate such that $n=0$ at the solid-liquid  interface. 
Substituting Eq.~\ref{c1} in Eq.~\ref{eq:quad}, we get $c_l=T_m$.
Differentiating Eq.~\ref{eq:quad} and setting $n=H_l$, 
\begin{equation}
    2 a_l H_l + b_l = \alpha 
    \label{eq:ab1}
\end{equation}
Eq.~\ref{c4} results in 
\begin{equation}
    a_l \frac{H_l^2}{3} + b_l \frac{H_l}{2}  = T_{avg} - T_m = \beta 
    \label{eq:ab2}
\end{equation}
Solving Eqns.~\ref{eq:ab1} and \ref{eq:ab2} for $a_l$ and $b_l$, 
\begin{equation}
    a_l = \frac{3}{2 H_l^2} \left( \frac{\alpha H_l}{2} - \beta \right)
\end{equation}
\begin{equation}
    b_l = \left( \frac{3\beta}{H_l} - \frac{\alpha}{2}\right)
\end{equation}
However, $\alpha$ is in terms of $T(H_l)$ as in Eq.~\ref{c3}, which is still unknown. Thus, we substitute the above expressions for $a_l$ and $b_l$ back in Eq.~\ref{eq:quad} and solve for $T(H_l)$. 
\begin{equation}
    T(H_l) = \frac{H_l\frac{h_a}{4k} T_a + \frac{3}{2} (T_{avg}-T_m) + T_m }{1+H_l\frac{h_a}{4k}}
\end{equation}
In the above equation, all terms on the RHS are known. Thus, the order of solving for the unknowns is as follows - $T(H_l)$, $\alpha$, $\beta$, $a_l$ and $b_l$. \\
A similar approach is adopted for the temperature profile in the solid region as well. We begin with assuming a temperature profile that varies linearly. 
\begin{equation}
    T(n) = a_s n + b_s
    \label{eq:linear}
\end{equation}
The constants $a_s$ and $b_s$ are determined by the conditions 
\begin{equation}
    T(H_s) = T_m
    \label{cs1}
\end{equation}
\begin{equation}
    \frac{\partial T}{\partial n} \Big|_{n=0} = \frac{h_s}{\kappa}(T(0)-T_s)
    \label{cs2}
\end{equation}
Substituting Eq.~\ref{cs1} in Eq.~\ref{eq:linear}, we get
\begin{equation}
    T_m = a_s H_s + b_s 
    \label{eq:cs1-1}
\end{equation}
Differentiating Eq.~\ref{eq:linear} and setting $n=0$,
\begin{equation}
    a_s = \frac{h_s}{\kappa}(T(0)-T_s)
\end{equation}
Substituting for $a$ in Eq.~\ref{eq:cs1-1}, 
\begin{equation}
 b_s = T_m - H_s  \frac{h_s}{\kappa}(T(0)-T_s)
\end{equation}
Since $T(0)$ is still unknown, we substitute $a_s$ and $b_s$ in Eq.~\ref{eq:linear} and solve for $T(0)$. 
\begin{equation}
    T(0) = \frac{T_m + \frac{h_s}{\kappa}T_s H_s}{ 1 +  \frac{h_s}{\kappa} H_s }
\end{equation}
We, therefore, solve for the unknowns in the following order - $T(0)$, $a_s$ and $b_s$.
\bibliographystyle{unsrt}

\bibliography{main.bib}
\end{document}